\documentclass[12pt]{article}
\usepackage{ascmac}
\usepackage{ulem}
\usepackage{latexsym}
\usepackage{graphicx}
\usepackage{amsmath}
\usepackage[top=3.5cm,bottom=3.5cm,left=2cm,right=2cm]{geometry}
\usepackage{pifont}
\usepackage{cancel}             
\usepackage{amsmath,amssymb}
\usepackage{color}

\newcommand{\h}{\hspace}
\newcommand{\be}{\begin{equation}}
\newcommand{\e}{\end{equation}}
\newcommand{\aln}[1]{\begin{align}#1\end{align}}

\begin{document}

\title{
\vbox{
\baselineskip 14pt
\hfill \hbox{\normalsize KUNS-2566
}} \vskip 1cm
\bf \Large Vanishing Higgs Potential in Minimal Dark Matter Models\vskip 0.5cm
}
\author{
Yuta~Hamada\thanks{E-mail:  \tt hamada@gauge.scphys.kyoto-u.ac.jp} and
Kiyoharu~Kawana\thanks{E-mail: \tt kiyokawa@gauge.scphys.kyoto-u.ac.jp} 
\bigskip\\
%\\*[20pt]
\it \normalsize
 Department of Physics, Kyoto University, Kyoto 606-8502, Japan\\
\smallskip
}
\date{\today}

\maketitle

\abstract{\normalsize

We consider the Standard Model with a new particle which is charged under $SU(2)_{L}$ with the hypercharge being zero. Such a particle is known as one of the dark matter (DM) candidates.
We examine the realization of the multiple point criticality principle (MPP) in this class of models.
Namely, we investigate whether the one-loop effective Higgs potential $V_{\text{eff}}(\phi)$ and its derivative $dV_{\text{eff}}(\phi)/d \phi$ can become simultaneously zero at around the string/Planck scale, based on the one/two-loop renormalization group equations. 
As a result, we find that only the $SU(2)_L$ triplet extensions can realize the MPP. More concretely, in the case of the triplet Majorana fermion, 
%$V_{\text{eff}}(\phi)=\frac{dV_{\text{eff}}(\phi)}{d \phi}=0$
the MPP is realized at the scale $\phi={\cal{O}}(10^{16}\text{GeV})$ if the top mass $M_{t}$ is around $172$ GeV. On the other hand, for the real triplet scalar, the MPP can be satisfied for $10^{16}\text{ GeV}\lesssim\phi\lesssim10^{17}$GeV and $172\text{ GeV}\gtrsim M_{t}\gtrsim171$ GeV, depending on the coupling between the Higgs and DM.

%We consider the multiple point \red{criticality} principle of the Standard Model (SM) with a new particle which is a $n_{\chi(X)}$ representation of $SU(2)_{L}$ with its hypercharge being zero. Such a model is well known as one of the \red{Dark Matter} candidates. We examine whether the one-loop effective Higgs potential $V_{\text{eff}}(\phi)$ and its derivative $dV_{\text{eff}}(\phi)/d \phi$ can become simultaneously zero, based on the two(one)-loop renormalization group equations. If we assume the perturbativity of theory up to the Planck/string scale, possibilities are \red{quite} limited: a new fermion (scalar) $\chi$ $(X)$ with $n_{\chi(X)}=3,5$ (3). Furthermore, from the point of view that the other vacuum should exist around the Planck/string scale, we can conclude that only $n_{\chi(X)}=3$ survives. %When $Y_{\chi}=0$,  it is known that the neutral component of $\chi$ can be a Dark Matter (DM) candidate, and its mass $M_{\chi}$ is determined by the cosmological relic abundance. As a result, the top mass $M_{t}$ and the scale $\Lambda_{\text{MPP}}$ at which the MPP is imposed are uniquely predicted. Furthermore, we find that $\Lambda_{\text{MPP}}$ is nicely approximated by the Landau Pole $\Lambda_{LP}$ of the $SU(2)_{L}$ gauge coupling $g_{2}$ when $n_{\chi}\geq6$. 
 }
\newpage

The discovery of the Higgs particle 
%and its mass 
\cite{Aad:2012tfa,Chatrchyan:2012ufa} is very meaningful for the Standard Model (SM).
The experimental value of the Higgs mass suggests that the Higgs potential can be stable up to the Planck scale $M_{pl}$ and also that both of the Higgs self coupling $\lambda$ and its beta function $\beta_{\lambda}$ become very small around $M_{pl}$. 
This fact attracts much attention, and there are many works which try to find its physical meaning~\cite{Froggatt:1995rt,Froggatt:2001pa,Meissner:2006zh,Foot:2007iy,Meissner:2007xv,Iso:2009ss,Iso:2009nw,Shaposhnikov:2009pv,Holthausen:2011aa,Bezrukov:2012sa,Hamada:2012bp,Iso:2012jn,Nielsen:2012pu,Jegerlehner:2013cta,Jegerlehner:2013nna,Hamada:2013cta,Buttazzo:2013uya,Branchina:2013jra,Kawamura:2013kua,Chao:2012mx,Kobakhidze:2014xda,Khan:2014kba,Khan:2015ipa,Spencer-Smith:2014woa,Haba:2014sia,Foot:2014ifa,Oda:2015kma}
 and implications for cosmology~\cite{Bezrukov:2007ep,Hamada:2013mya,Jegerlehner:2014mua,Hamada:2014iga,Fairbairn:2014zia,Bezrukov:2014bra,Enqvist:2014bua,Hook:2014uia,Haba:2014zda,Hamada:2014xka,Ko:2014eia,Haba:2014zja,He:2014ora,Herranen:2014cua,Hamada:2014wna,Hamada:2014raa,Hamada:2015ria,Okada:2015zfa,Inagaki:2015fva,Jegerlehner:2015cva,Abe:2015bba,DiLuzio:2015oha,Bamba:2015uxa,Nurmi:2015ema,Sebastiani:2015kfa,Herranen:2015ima}.

%%%%%%%%%%%%%%%%%%
In \cite{Froggatt:1995rt,Froggatt:2001pa}, 
the Higgs mass was predicted
%it was argued that the Higgs mass can be predicted
 to be around $130$GeV by the requirement that %the running Higgs self coupling
 $\lambda(\mu)$ and $\beta_{\lambda}(\mu)$ simultaneously become zero around $M_{pl}$.\footnote{
It is interesting that the quadratic divergent bare Higgs mass also vanishes around this scale~\cite{Hamada:2012bp}.} 
Namely, the minimum of the Higgs potential $V(\phi)$ around $M_{pl}$ vanishes.
Such a requirement
% (not always at $M_{pl}$) 
is called the multiple point criticality principle (MPP), and there have been many suggestions \cite{Kawai:2011qb,Kawai:2013wwa,Hamada:2014ofa,Hamada:2014xra,Hamada:2015wea,Hamada:2015dja,Hamada:2015ria,Hamada:2014xka,Kawana:2014zxa,Kawana:2015tka} that this principle might be closely related to physics at the Planck scale.
One of the good points of the principle is its predictability: The low-energy effective couplings are fixed so that the minimum of the potential takes zero around $M_{pl}$.
See \cite{Hamada:2014xka,Kawana:2014zxa,Okada:2014nea,Kawana:2015tka,Haba:2015rha} for examples of the prediction.\\
 
%%%%%%%%%%%%%%%%%%
By taking the fact that the MPP is realized in the SM into consideration, a natural question is whether the MPP can be also realized in the models beyond the SM. 
It is meaningful to consider the MPP of these models because we can understand whether the SM is actually special among them.
One of the interesting extensions is adding a new weakly interacting fermion $\chi$ or scalar $X$, which is a $n_{\chi(X)}$ representation of $SU(2)_{L}$  with the hypercharge $Y_{\chi(X)}$. 
Such extensions are phenomenologically well studied because they have dark matter (DM) candidates when $Y_{\chi(X)}=0$ \cite{Cirelli:2005uq,Hisano:2006nn,Cirelli:2007xd}. 
In this paper, we focus on $Y_{\chi(X)}=0$, that is, Majorana fermions and real scalars. We examine the realization of the MPP of these models% of these extensions 
, based on the one/two-loop renormalization group equations (RGEs). 
We use the effective Higgs self coupling $\lambda_{\text{eff}}$ and its beta function $\beta_{\lambda_{\text{eff}}}$ defined from the one-loop effective Higgs potential $V_{\text{eff}}(\phi)$. Their definitions and the two-loop RGEs when we add a new fermion are presented in \ref{app:rge}. 
%
%When a new particle is a scalar, we have to pay attention to the meaning of the MPP because the whole potential is a function of two scalar fields\blue{(MPPがwhole potentialがvanishすることだとは言っていない)}. 
%
%Generally speaking, it is difficult to make all scalar couplings and their beta functions simultaneously zero (see \cite{Hamada:2014xka,Kawana:2014zxa} for examples). 
%
In the case of the new scalar (fermion), we only have to consider $n_{X}=3$ ($n_{\chi}=3,5$) since the scalar couplings ($SU(2)_{L}$ coupling $g_{2}$) rapidly blow(s) up when $n_{X}\geq4$  \cite{Hamada:2015bra} ($n_{\chi}\geq7$ \cite{Cirelli:2005uq}), and the theory does not valid up to $M_{pl}$. For the septet and nonet fermion cases, we discuss this point in \ref{app:LP}.
%%%%%%%%%%%%%%%

In the following discussion, we regard the top mass $M_{t}$ as a free parameter, and the Higgs mass is varied within \cite{Aad:2015zhl}
%%%%%%%%%%%%%%%
\be M_{h}=125.09\pm0.32\text{GeV}.\e
%%%%%%%%%%%%%%%
%\red{As we will see in the following discussion, our results are not sensitive to the above values of the Higgs mass. Namely, the predicted values of $M_{t}$ (the scale $\Lambda_{\text{MPP}}$ at which the minimum of the Higgs potential vanishes) changes at most by ${\cal{O}}(0.5\text{GeV})$ (twice). }
%
As for the initial values of the $\overline{\text{MS}}$ SM couplings,
we use the results of \cite{Buttazzo:2013uya}. 
For illustration, the $Y_{\chi}\neq0$ cases are also discussed in \ref{app:rem}. \\
 
%%%%%%%%%%%%%%%
First, we consider a new fermion. For $n_{\chi}=3$ and $5$, the mass $M_{\chi}$ is determined by the thermal relic abundance \cite{Hisano:2006nn,Cirelli:2007xd}:
\be M_{\chi}\simeq\begin{cases}2.8\h{1mm}\text{TeV}&(\text{for $n_{\chi}=3$}),\\
10\h{1mm}\text{TeV}&(\text{for $n_{\chi}=5$}).
\end{cases}\label{eq:DMmass}\e
As a result, $M_{t}$ and $\Lambda_{\text{MPP}}$ are uniquely predicted because there is no additional free parameter. %\footnote{In fact, there are additional Yukawa couplings when $n_{\chi}=3$ (Type III seesaw): $y_{\chi}^{i}\bar{L}_{i}\chi H^{\dagger}$.
% 
%\red{However, because $M_{\chi}=2.8$ TeV is so small, this effect can be neglected if we assume that the small neutrino masses are generated by the seesaw mechanism.}}. 
%
The results are
\aln{&171.7\text{GeV}\leq M_{t}\leq172.0\text{GeV}\h{2mm},\h{2mm} 2.5\times10^{16}\text{GeV}\leq\Lambda_{\text{MPP}}\leq3.2\times10^{16}\text{GeV} \h{2mm}(\text{for $n_{\chi}=3$}),\nonumber\\
&174.8\text{GeV}\leq M_{t}\leq175.2\text{GeV}\h{2mm},\h{2mm}1.1\times10^{11}\text{GeV}\leq\Lambda_{\text{MPP}}\leq1.2\times10^{11}\text{GeV}\h{2mm}(\text{for $n_{\chi}=5$}),\label{eq:MPPpre}}
depending on $124.77\text{GeV}\leq M_{h}\leq125.41\text{GeV}$.\footnote{These values of $M_{t}$ are consistent with the recent analyses: $M_{t}=173.34\pm0.76$GeV \cite{ATLAS:2014wva} and $M_{t}=172.38\pm 0.10\pm0.65$GeV \cite{CMS:2014hta} at $2\sigma$ level. However, the relation between these masses and the pole mass is not clear. In the following calculation of the bare Higgs mass, we use more conservative value of $M_{t}$ determined by the $t\bar{t}$ total cross section \cite{Moch:2014tta}.}
%\be \left(M_{t},\Lambda_{\text{MPP}}\right)=\begin{cases}\left(176.6,\h{1mm}6.8\times10^{7}\right)\text{GeV}&\text{for 7-plet, $M_{DM}=10$TeV}\\
%\left(174.9,\h{1mm}1.0\times10^{9}\right)\text{GeV}&\text{for 7-plet, $M_{DM}=100$TeV}\end{cases}.\e
%\be \left(M_{t},\Lambda_{\text{MPP}}\right)=\begin{cases}\left(165.2,\h{1mm}3.0\times10^{5}\right)\text{GeV}&\text{for 9-plet, $M_{DM}=10$TeV}\\
%\left(162.2,\h{1mm}3.4\times10^{6}\right)\text{GeV}&\text{for 9-plet, $M_{DM}=100$TeV}\end{cases}.\e
%%%%%%%%%%%%%%%%%%%%%%%%%%%%%%PLOT1%%%%%%%%%%%%%%%%%%%%%%%%%%%%%%%%%%
\begin{figure}
\begin{center}
\begin{tabular}{c}
\begin{minipage}{0.5\hsize}
\begin{center}
\includegraphics[width=9cm]{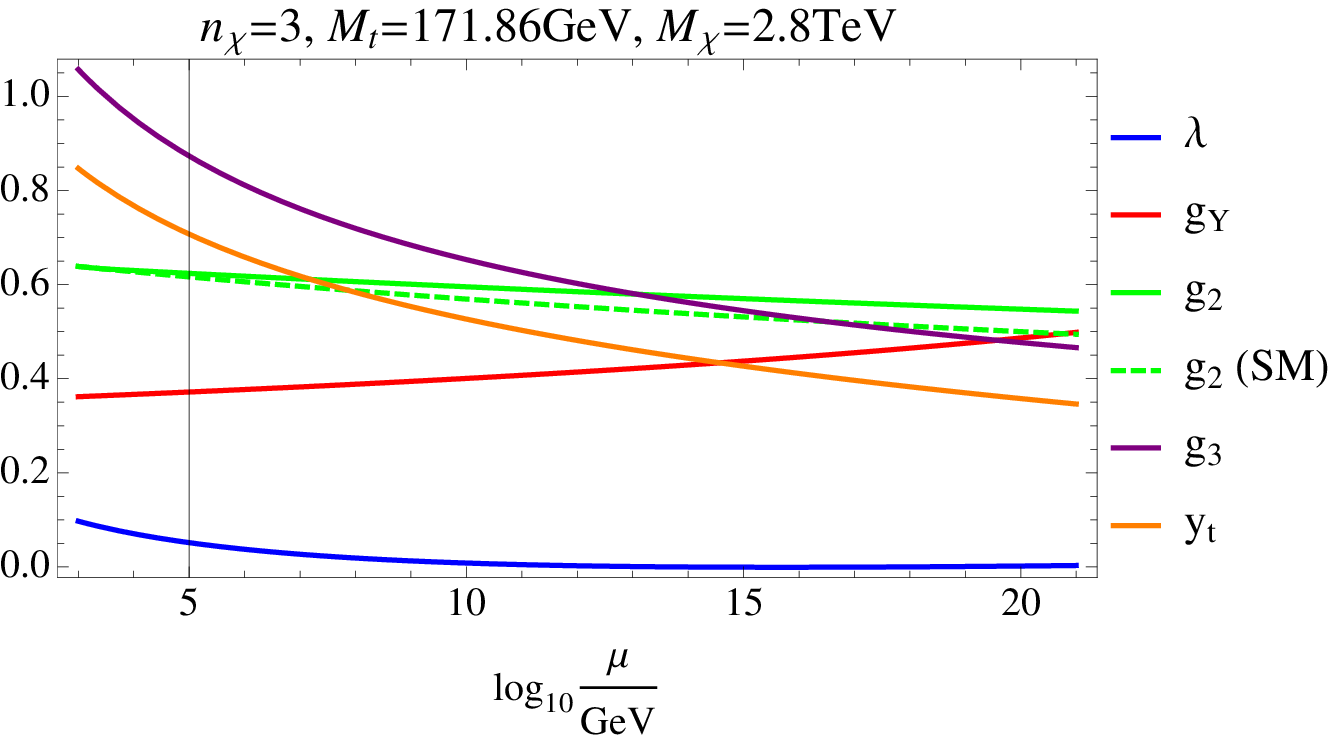}
\end{center}
\end{minipage}
\begin{minipage}{0.5\hsize}
\begin{center}
\includegraphics[width=9cm]{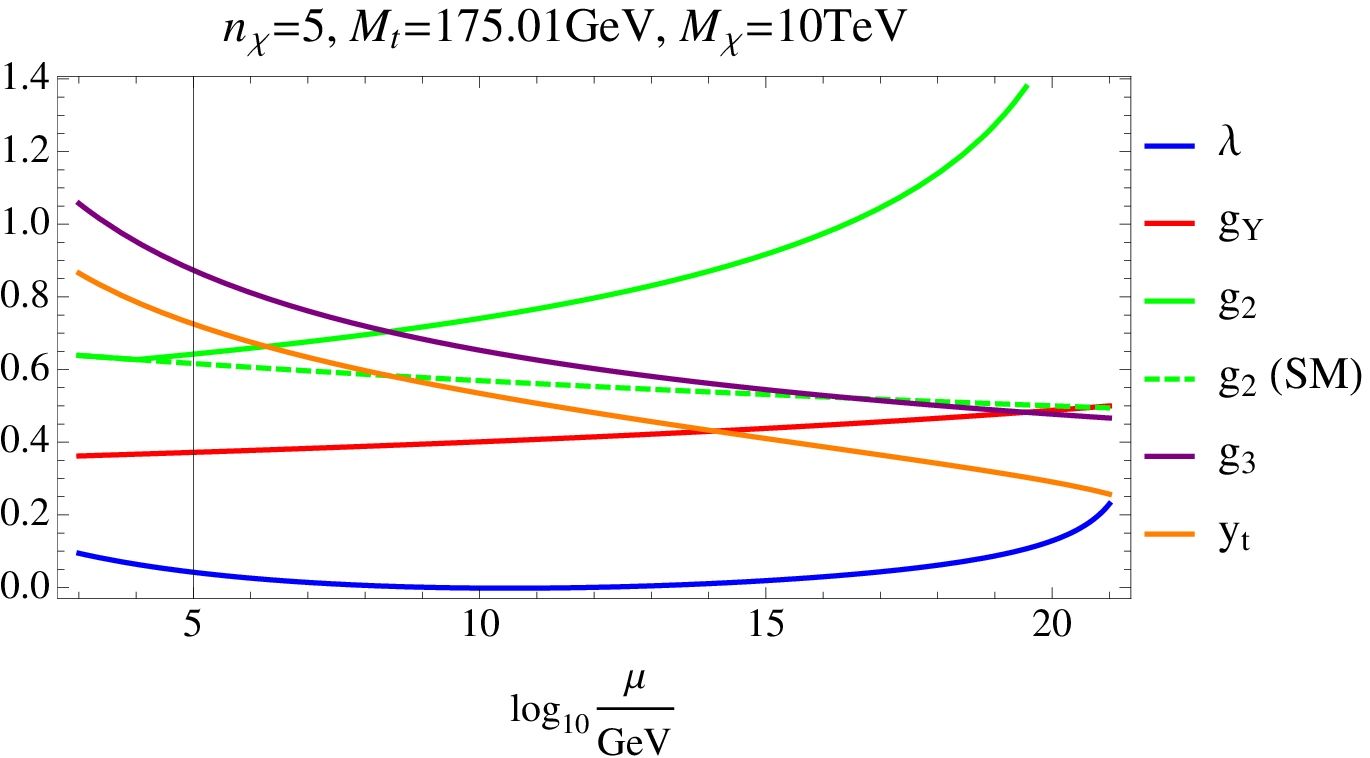}
\end{center}
\end{minipage}
\\
\\
\begin{minipage}{0.5\hsize}
\begin{center}
\includegraphics[width=8.5cm]{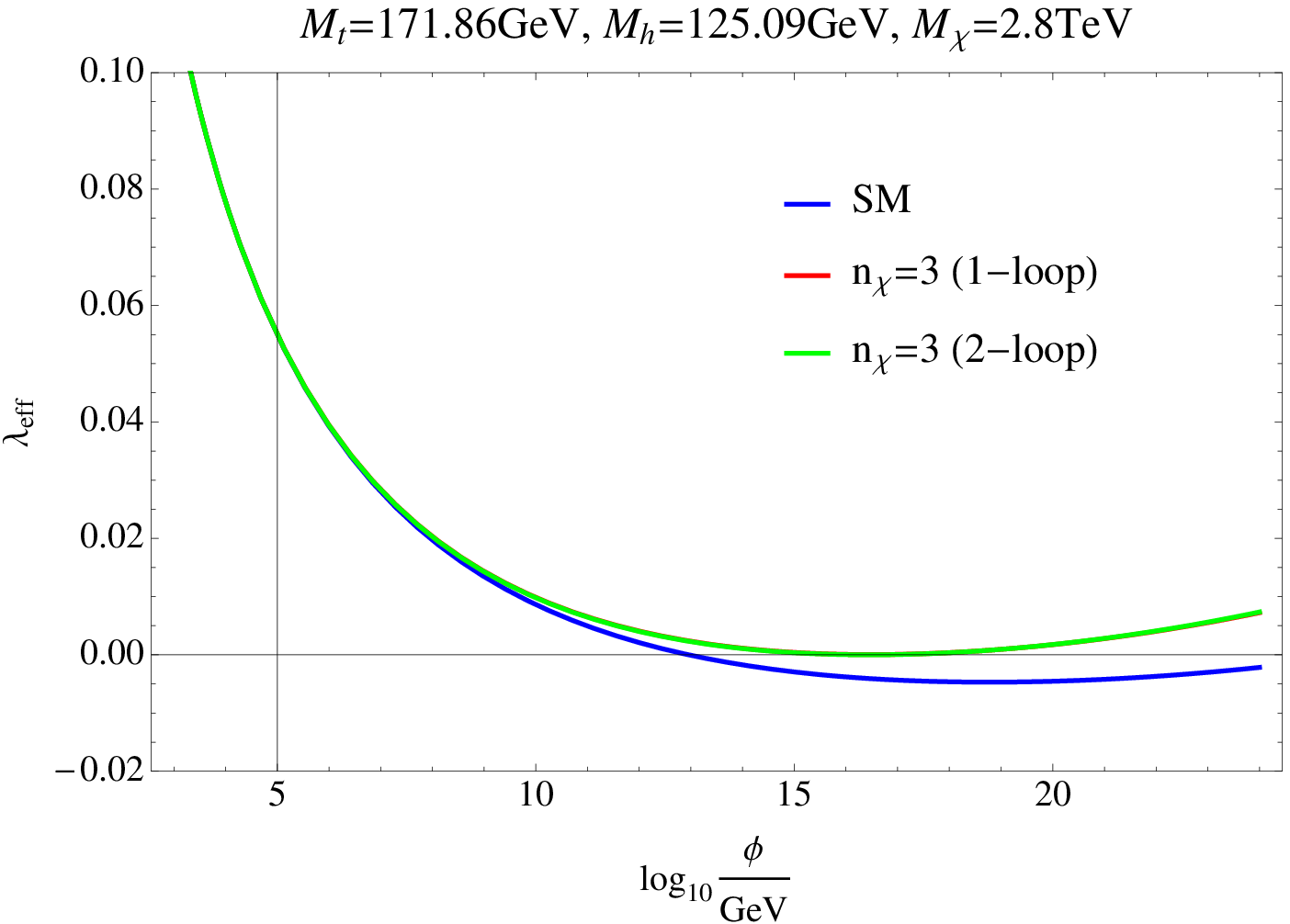}
\end{center}
\end{minipage}
\begin{minipage}{0.5\hsize}
\begin{center}
\includegraphics[width=8.5cm]{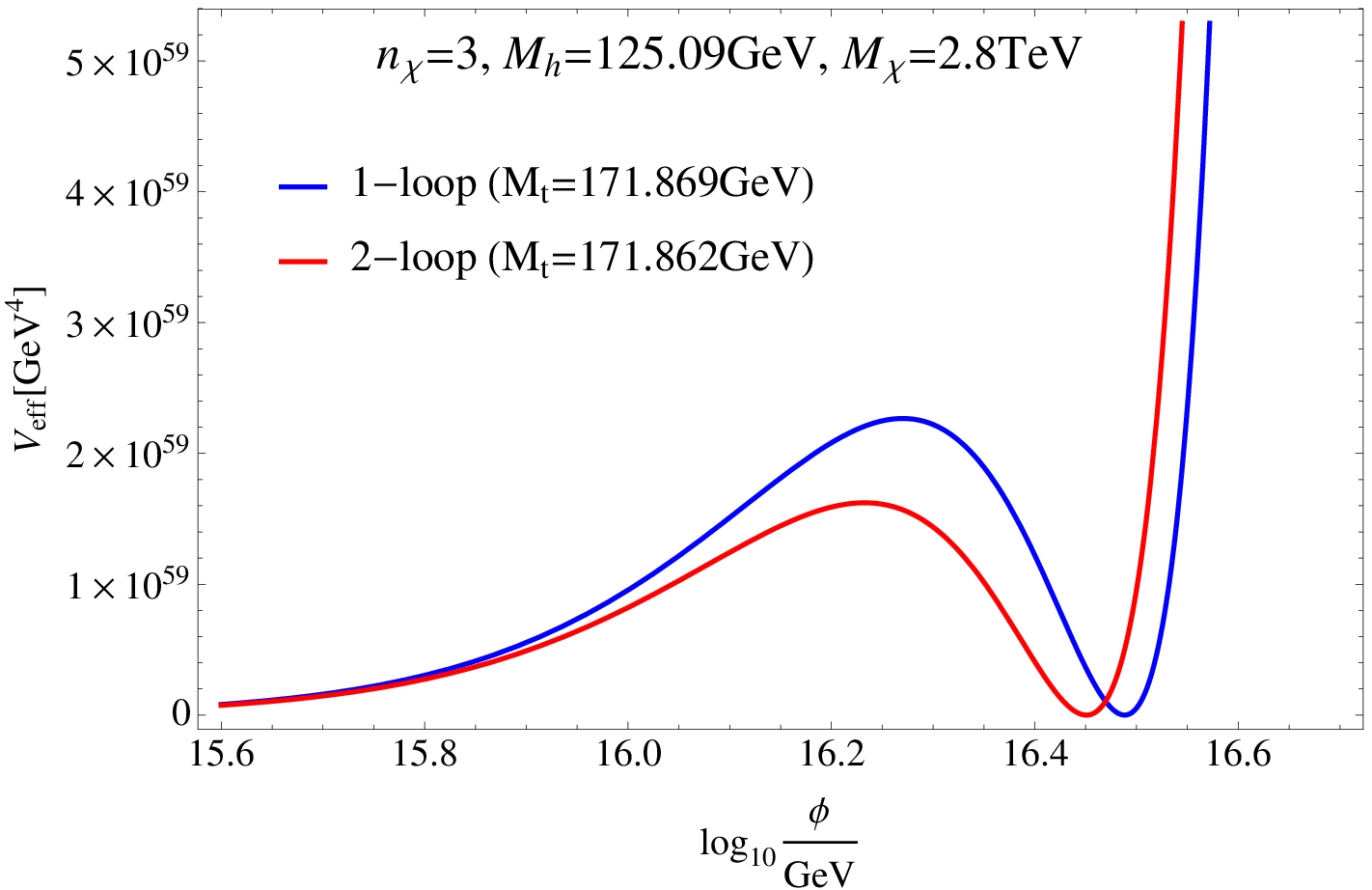}
\end{center}
\end{minipage}
\\
\\
\begin{minipage}{0.5\hsize}
\begin{center}
\includegraphics[width=8cm]{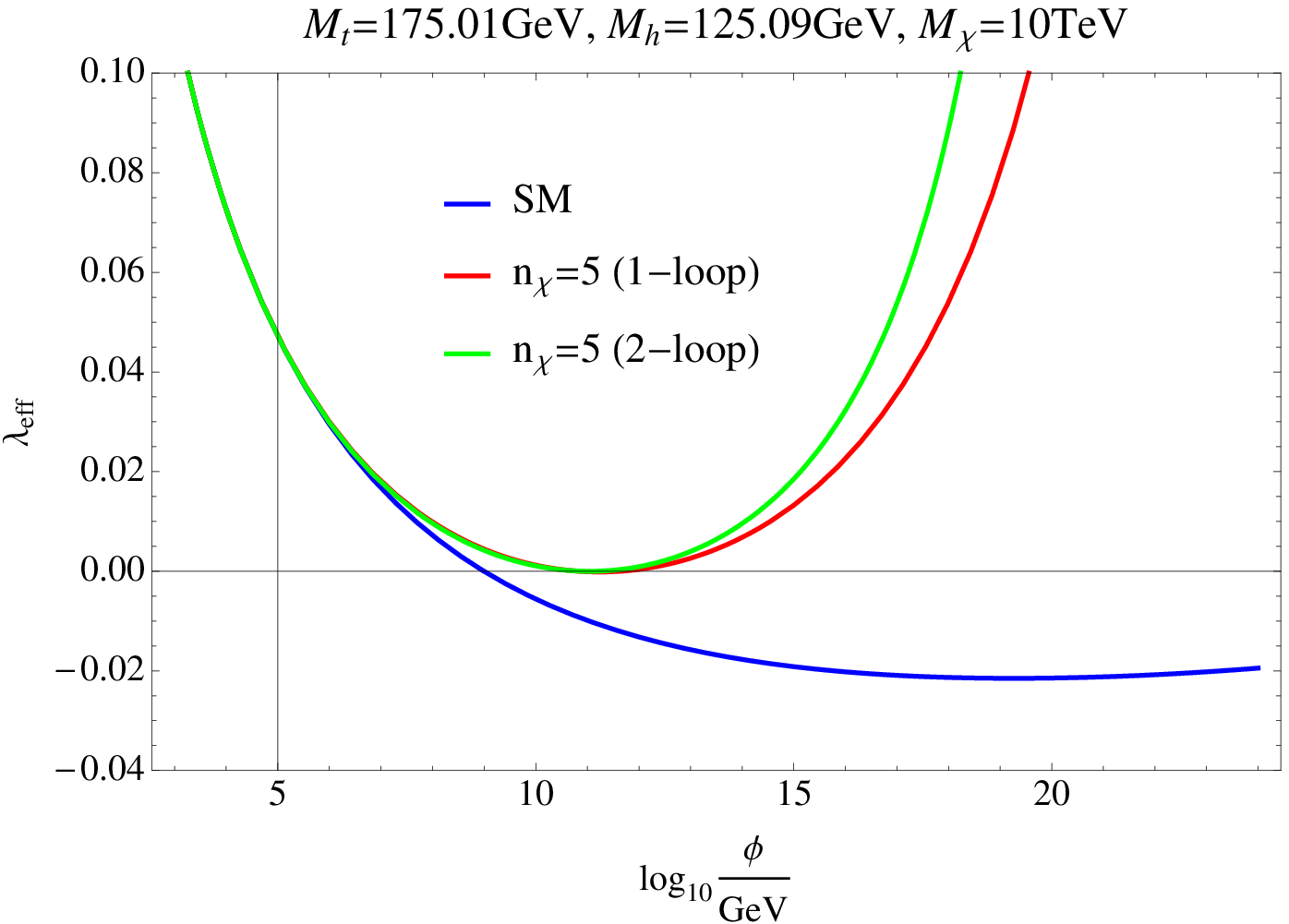}
\end{center}
\end{minipage}
\begin{minipage}{0.5\hsize}
\begin{center}
\includegraphics[width=8.5cm]{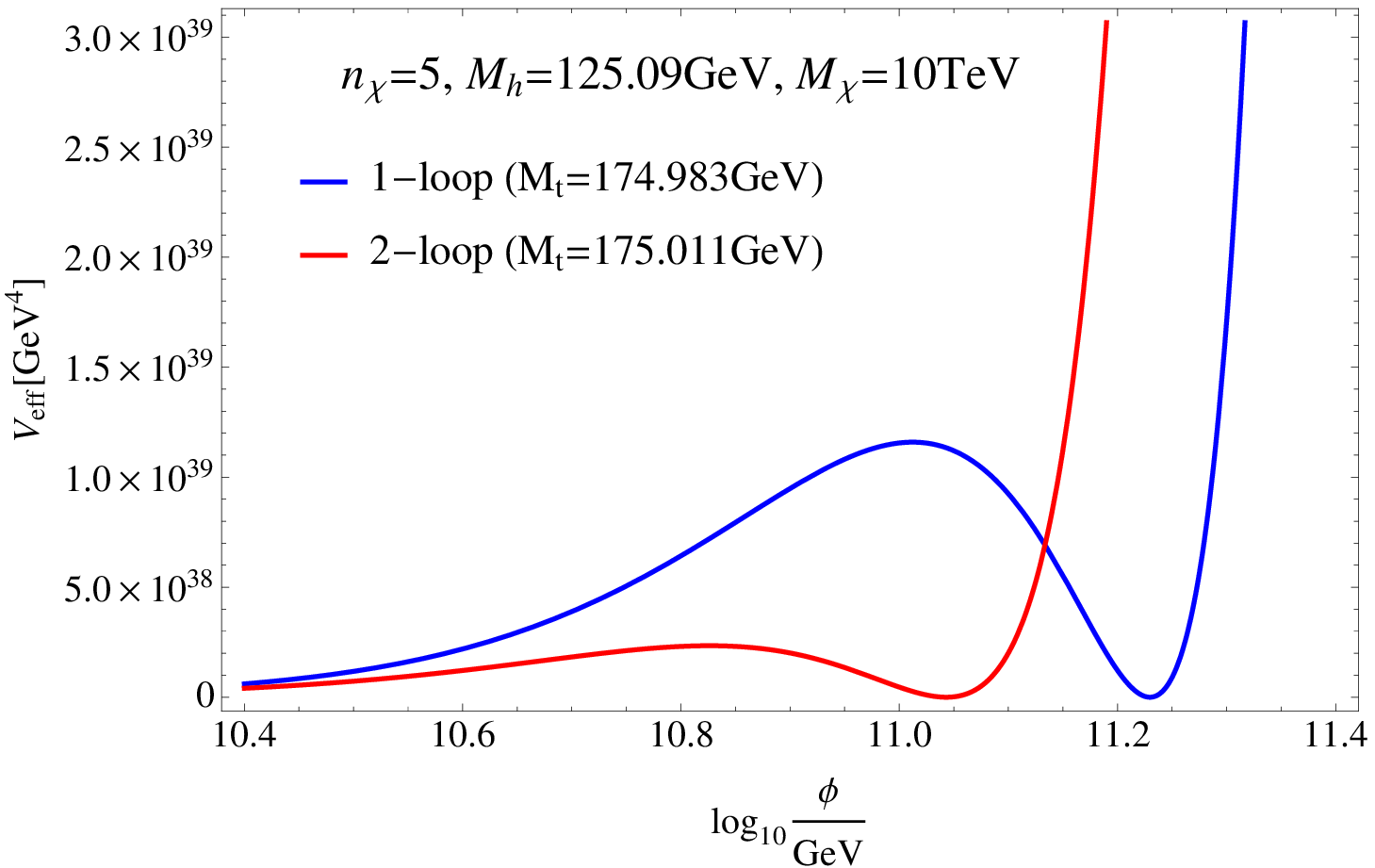}
\end{center}
\end{minipage}
\end{tabular}
\end{center}
\caption{Upper left (right): the runnings of the SM parameters when $n_{\chi}=3$ $(5)$. Here, the dashed green lines represent the SM running of $g_{2}$. Middle (Lower): the running of the effective Higgs self coupling $\lambda_{\text{eff}}$ (left) and the one-loop effective Higgs potential $V_{\text{eff}}(\phi)$ (right) in the case of $n_{\chi}=3$ (5).}
\label{fig:DM}
\end{figure}
%%%%%%%%%%%%%%%%%%%%%%%%%%%%%%%%%%%%%%%%%%%%%%%%%%%%%%%%%%%%%%%%%%%%
The upper panels of  Fig.\ref{fig:DM} show the runnings of the SM parameters where $M_{h}=125.09$GeV, and $M_{t}$ is correspondingly fixed so that the MPP is realized.
Here, we also show the SM running of $g_{2}$ by the dashed green line for comparison.
Furthermore, in the middle and lower panels, we show the corresponding $\lambda_{\text{eff}}$ (left) and $V_{\text{eff}}(\phi)$ (right). 
In these figures, the one-loop results are also shown.
One can actually see that the potential and its derivative simultaneously become zero at a high energy scale, and that the only triplet can have the other vacuum near the string/Planck scale. We note that the two-loop effects are small. \\

%%%%%%%%%%%%%%%%%%%%%%%%%SCALAR%%%%%%%%%%%%%%%%%%%%%%%%%%%%%
\begin{figure}
\begin{center}
\begin{tabular}{c}
\begin{minipage}{0.5\hsize}
\begin{center}
\includegraphics[width=8.5cm]{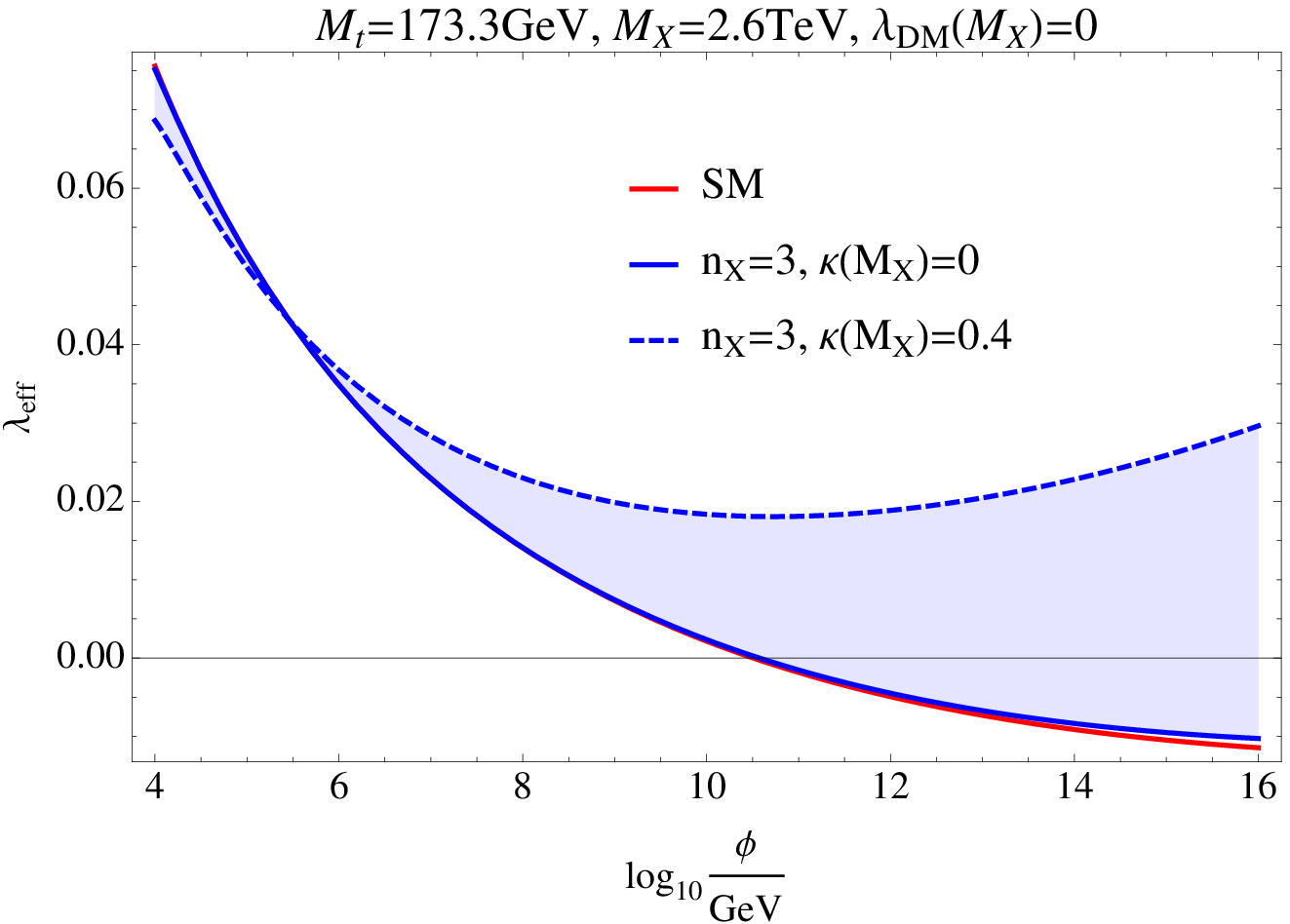}
\end{center}
\end{minipage}
\begin{minipage}{0.5\hsize}
\begin{center}
\includegraphics[width=8.5cm]{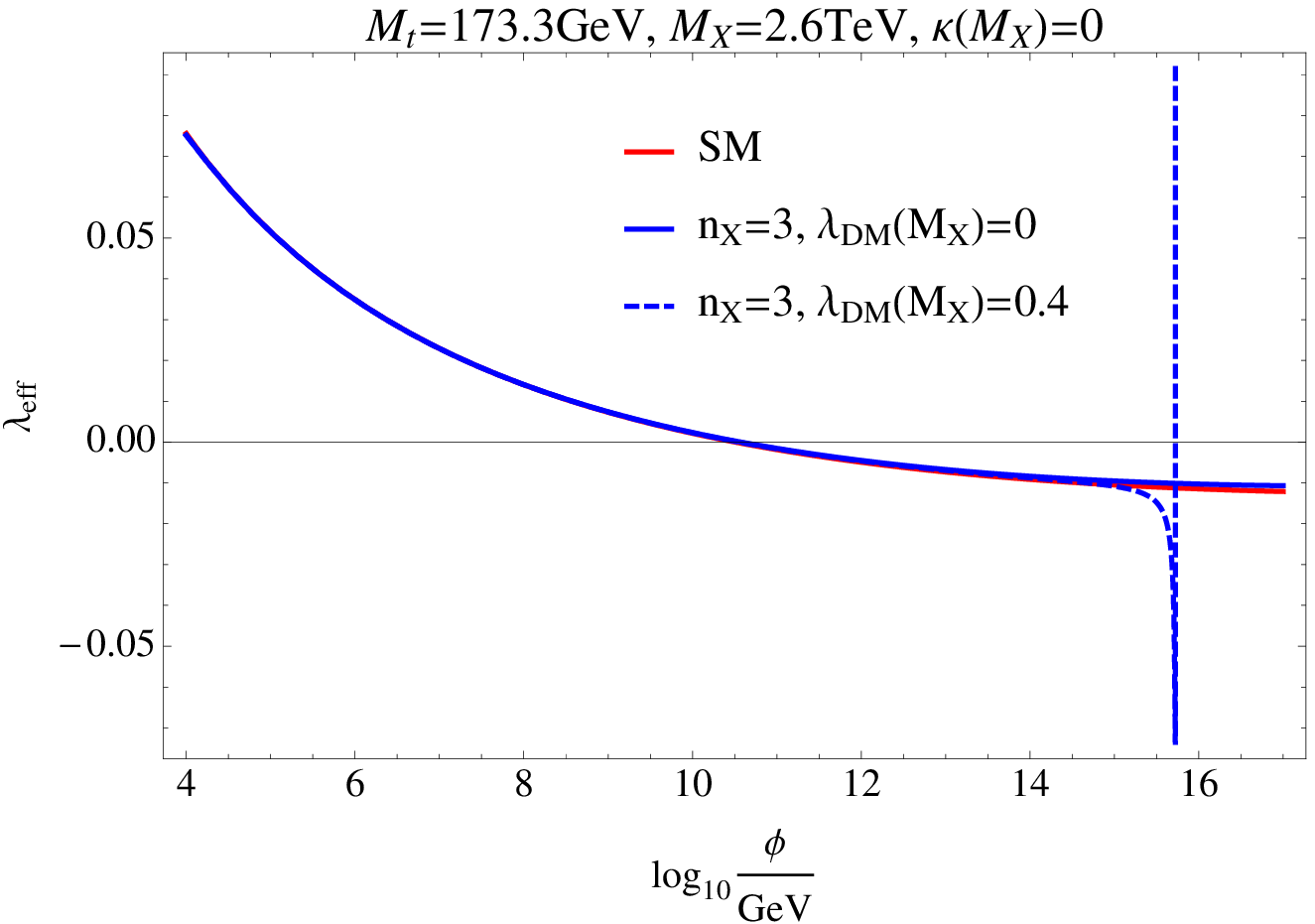}
\end{center}
\end{minipage}
\\
\\
\begin{minipage}{0.5\hsize}
\begin{center}
\includegraphics[width=8.5cm]{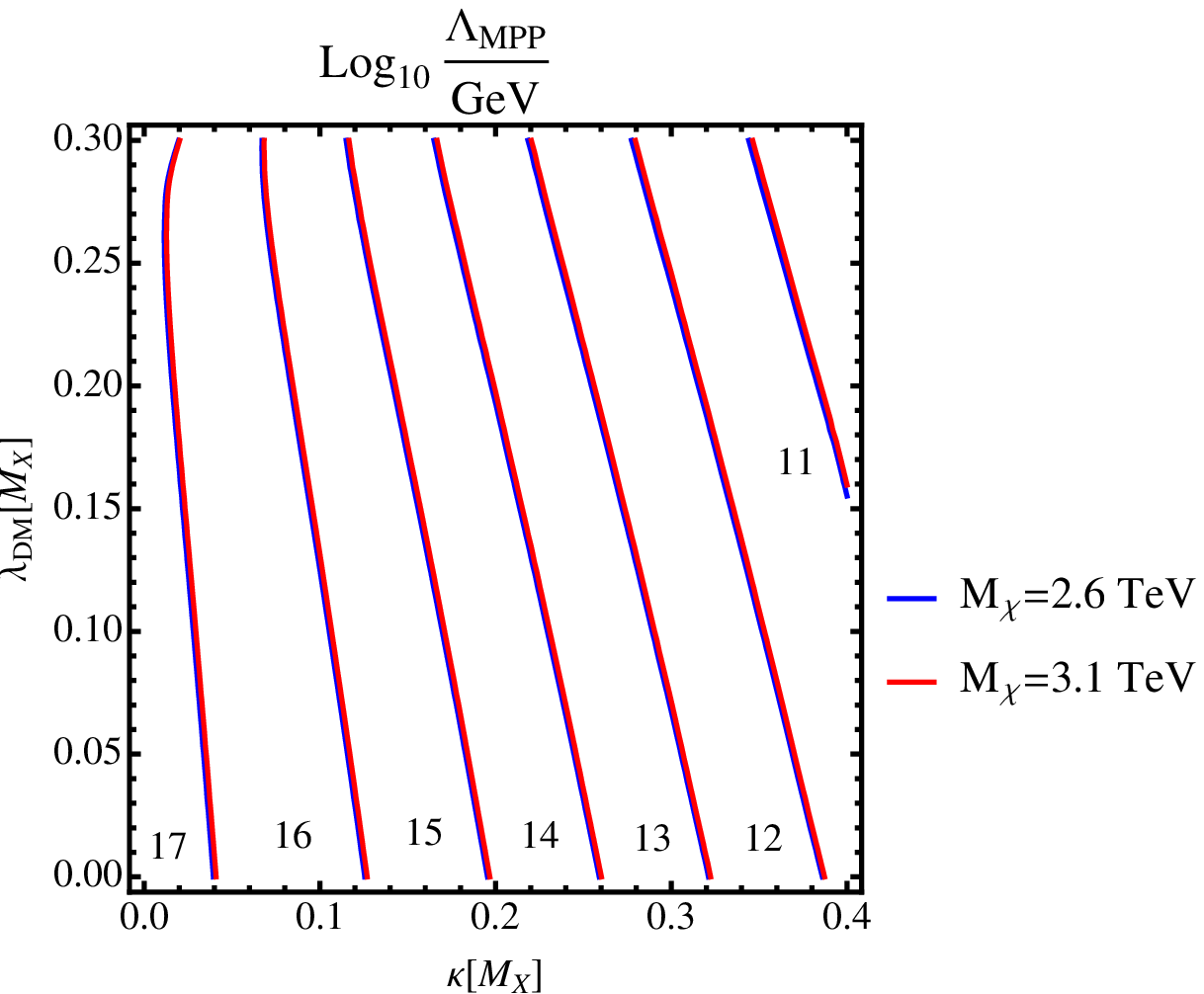}
\end{center}
\end{minipage}
\begin{minipage}{0.5\hsize}
\begin{center}
\includegraphics[width=8.5cm]{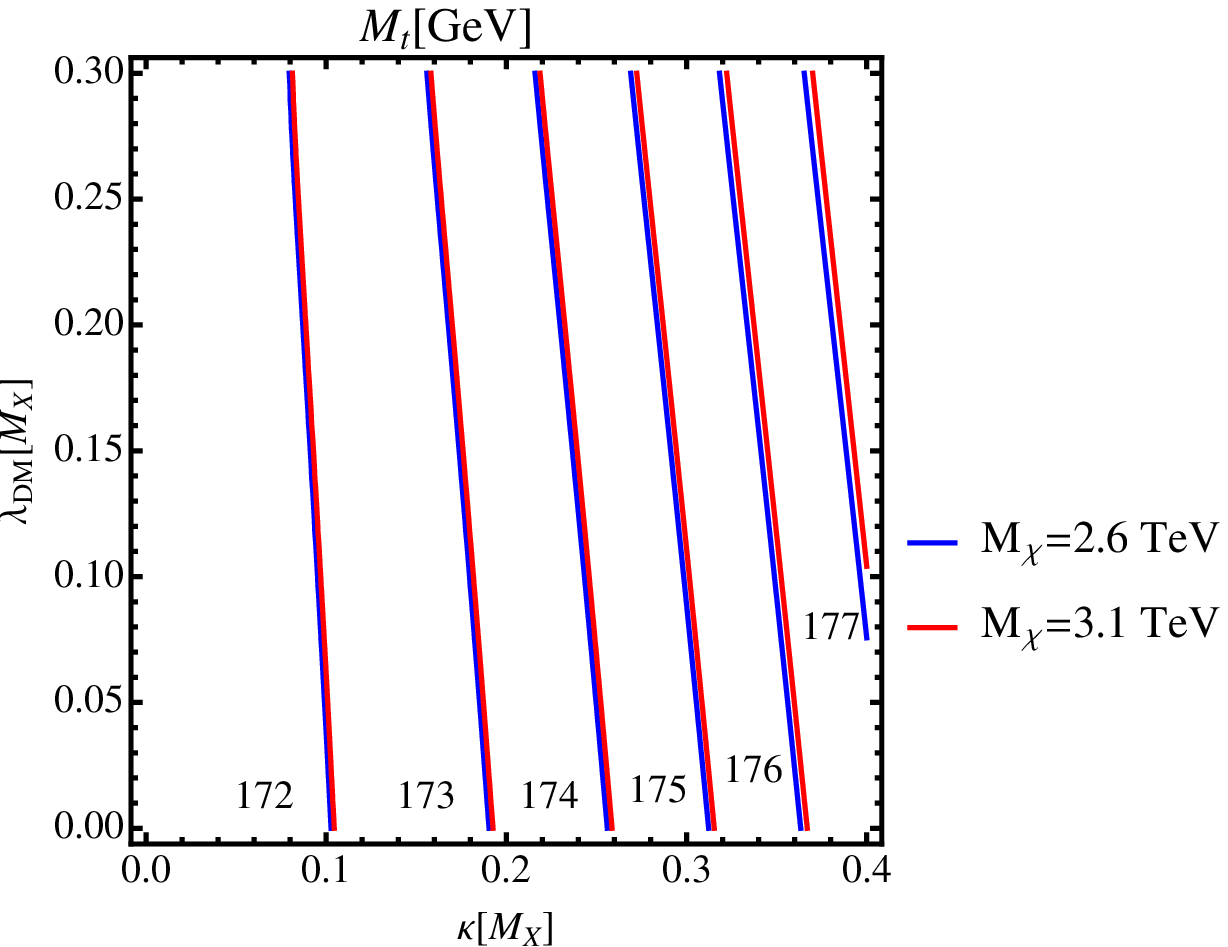}
\end{center}
\end{minipage}
\end{tabular}
\end{center}
\caption{Upper: the running of $\lambda_{\text{eff}}$ in the case of $n_{X}=3$. Here, the blue band of the left panel corresponds to the change of $\kappa$ at $\mu=M_{X}$ from 0 to $0.4$. Lower: $\Lambda_{\text{MPP}}$ (left) and $M_{t}$ (right) as a function of $\kappa$ and $\lambda_{DM}$ at $\mu=M_{X}$. The blue (red) contours correspond to $M_{X}=2.6\h{1mm}(3.1)$ TeV.}
\label{fig:MPPs}
\end{figure}

Now let us consider a new scalar.  As mentioned before, the remaining possibility is $n_{X}=3$~\cite{Hamada:2015bra}.
The potential of the scalar fields is 
\be V=-\frac{M_{h}^{2}}{2}H^{\dagger}H+\frac{M_{X}^{2}}{2}XX+\lambda\left(H^{\dagger}H\right)^{2}+\lambda_{DM}\left(XX\right)^{2}+\kappa\left(H^{\dagger}H\right)\left(XX\right).\e
Here, $H$ is the SM Higgs doublet. The one-loop RGEs which are different from those of the SM are as follows\footnote{As one can see from the results of the fermion cases, the two-loop effects are small when we consider the MPP. This is why we consider the one-loop beta functions here. }:
\begin{align}
\frac{dg_{2}}{dt}&=-\frac{g_{2}^{3}}{(4\pi)^{2}}\frac{17}{6},\\
\frac{d\lambda}{dt}&=\frac{1}{16\pi^{2}}\Bigg(\lambda\left(24
   \lambda-9
   g_{2}^{2}-3  g_{Y}^{2}+12
   y_{t}^{2}\right)+\frac{3}{2}\kappa^{2}+\frac{3}{4}g_{Y}^{2}g_{2}^{2}+\frac{9}{8} g_{2}^{4}+\frac{3}{8}
   g_{Y}^{4}-6 y_{t}^{4}\Bigg),\\
\frac{d\lambda_{DM}}{dt}&=\frac{1}{16\pi^{2}}\left(22\lambda_{DM} ^{2}+2\kappa^{2}-24g_{2}^{2}\lambda_{DM}+12g_{2}^{4}\right),\\
\frac{d\kappa}{dt}&=\frac{1}{16\pi^{2}}\Bigg(4\kappa^{2}+12\kappa\lambda+10\kappa\lambda_{DM}+6y_{t}^{2}\kappa-\frac{33}{2}g_{2}^{2}\kappa-\frac{3}{2}g_{Y}^{2}\kappa+6g_{2}^{4}\Bigg).
\end{align}
Furthermore, there is an additional contribution to $V_{\text{eff}}(\phi)$:
\be \Delta V_{\text{1loop}}(\phi)=\frac{3m_{DM}(\phi)^{4}}{64\pi^{2}}\left(\ln\left(\frac{m_{DM}(\phi)^{2}}{\phi^{2}}\right)-\frac{3}{2}\right),\e
where 
\be m_{DM}(\phi)=\sqrt{M_{X}^{2}+\kappa(\phi)e^{2\Gamma(\phi)}\phi^{2}}.\label{eq:DMmasspot}\e
In this case, the thermal abundance of $X$ depends on the value of $\kappa$.
%Although $M_{X}$ \blue{is fixed by} the thermal relic abundance, however, it depends on \blue{the value of }$\kappa$ in this case.
% 
Here we use 
\be M_{X}=2.6\h{2mm}\text{TeV and }3.1\h{2mm}\text{TeV}\label{eq:darkmass}\e 
for our calculation \footnote{The mass of a new scalar suffers from fine-tuning problem. However, because our motivation in this paper is to distinguish the minimal dark matter models in the context of the MPP, we take Eq.(\ref{eq:darkmass}) as the dark matter mass.}. 
$M_{X}=2.6$ TeV and $M_{X}=3.1$ TeV correspond to $\kappa=0$ and $\kappa=1$, respectively~\cite{Cirelli:2007xd}. 
The upper panels of Fig.\ref{fig:MPPs} show the runnings of $ \lambda_{\text{eff}}$ when $M_{X}=2.6$ TeV.  
Here, the blue band of the left panel corresponds to the change of $\kappa$ at $\mu=M_{X}$ from 0 to $0.4$. In the case of $\lambda_{DM}(M_{X})=0.4$ of  the right panel, the rapid increase of $\lambda_{\text{eff}}$ around $10^{16}$ GeV is due to the Landau pole of $\lambda_{DM}$. Namely, $\lambda_{DM}$ becomes infinity below $M_{pl}$.
The lower left (right) panel of Fig.3 shows the contour plot of $\Lambda_{\text{MPP}}\h{1mm}(M_{t})$ as a function of $\lambda_{DM}$ and $\kappa$ at $\mu=M_{X}$. The blue (red) contours correspond to $M_{X}=2.6\h{1mm}(3.1)$ TeV.
One can see that $\Lambda_{\text{MPP}}$ is close to the string/Planck scale when $\kappa(M_{X})\lesssim0.1$ and $M_{t}\lesssim172$GeV.  \\

\begin{figure}
\begin{center}
\begin{tabular}{c}
\begin{minipage}{0.5\hsize}
\begin{center}
\includegraphics[width=8.5cm]{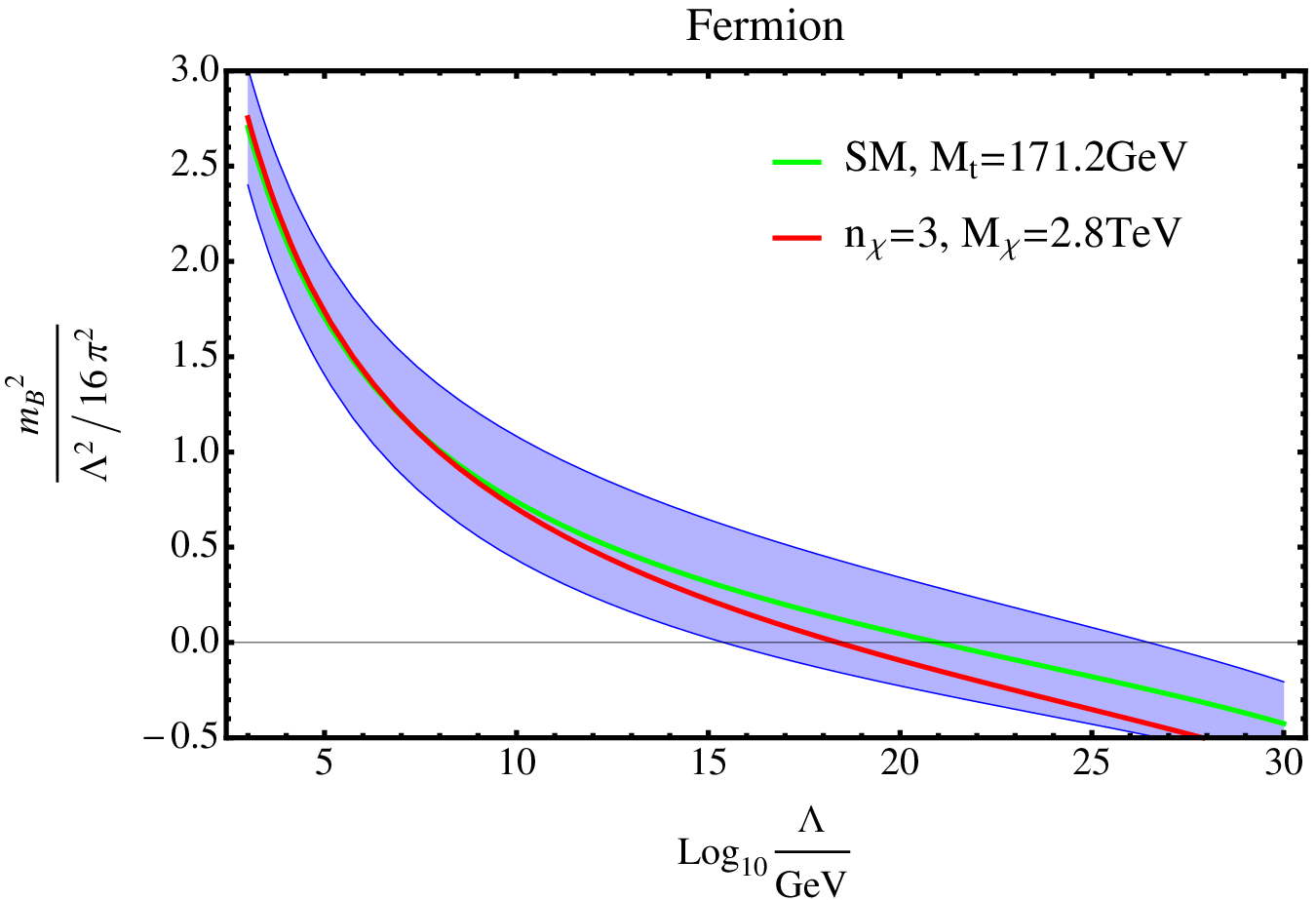}
\end{center}
\end{minipage}
\begin{minipage}{0.5\hsize}
\begin{center}
\includegraphics[width=8.5cm]{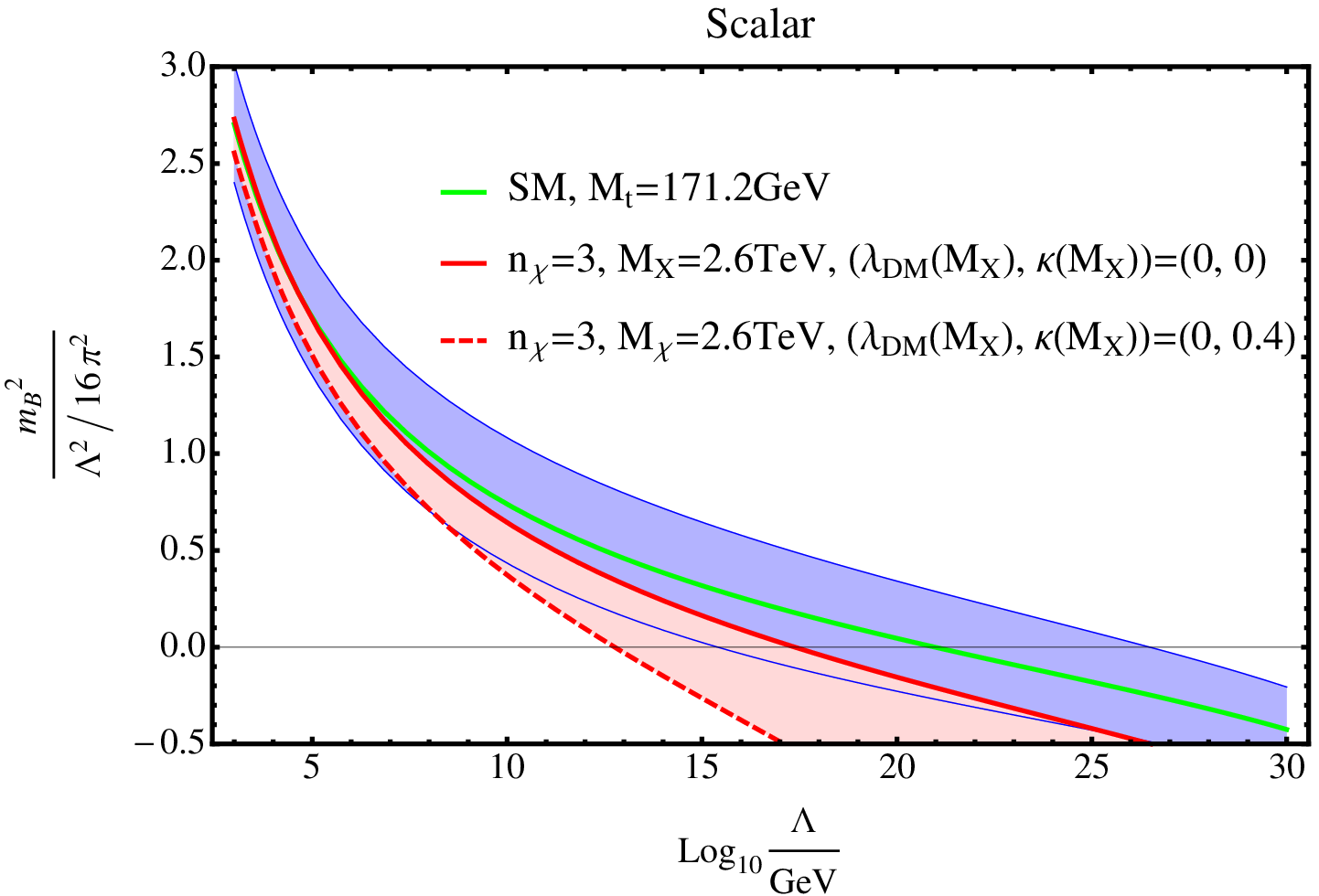}
\end{center}
\end{minipage}
\end{tabular}
\end{center}
\caption{The bare Higgs mass $\frac{m_{B}^{2}}{16\pi^{2}\Lambda^{2}}$ as a function of a cut-off scale $\Lambda$. Here the blue bands (red band) correspond(s) to the $2\sigma$ deviation from $M_{t}=171.2$ GeV (the change of $\kappa$ at $\mu=M_{X}$ from 0 to $0.4$). }
\label{fig:bare}
\end{figure}

%%%%%%%%%%%%%%%%%%%
In order to discuss the Higgs potential around the cutoff scale $\Lambda$,
%In addition to the MPP, 
it is meaningful to consider how the existence of a new particle changes the behavior of the bare Higgs mass $m_{B}$ as a function of $\Lambda$.\footnote{
Within field theory, the quadratic divergence does not appear after the renormalization.
However, it can have the physical meaning if we consider the scale around the Planck/string one, because the SM couples with the gravity.
In this paper, we assume that the physics around the Planck scale is described by string theory, which is the cutoff theory whose universal cutoff scale is given by the string scale.
%In string theory, the universal cutoff is given by the string scale.
This is why we take the universal cutoff in the calculation of $m_B^2$.
%We assume that the cutoff theory  
%universal cutoff .
See \cite{Hamada:2012bp} for the detail. }
This is because $m_{B}$ would appear in the Higgs potential above $\Lambda$~\cite{Hamada:2013mya}.  
%and it is related to the 
We now examine whether $m_{B}$ vanishes around the string scale or not.\footnote{The vanishing bare mass is so-called Veltman condition~\cite{Veltman:1980mj}. From the point of view of low energy field theory, $m_{B}^2=0$ is accidental and seems to require the fine-tuning at the Planck scale. We hope that $m_{B}^2=0$ comes from some mechanism related to the physics at the Planck scale. }
See \cite{Hamada:2012bp} for the evaluation of $m_{B}$ in the SM.

Here, let us focus on $(n_{\chi(X)},Y_{\chi(X)})=(3,0)$ at one-loop level. 
For $\chi$, $m_B$ is given by 

%We do not follow detail calculations because they are already presented in \cite{Hamada:2012bp}. 
\begin{align}
\frac{m_{B}^{2}|_{\text{1-loop}}}{\Lambda^{2}/16\pi^{2}}
=-\left(
6\lambda+{3\over4}g_Y^2+{9\over4}g_2^2-6y_t^2
\right),
\end{align}
where the couplings are evaluated at $\mu=\Lambda$.
%For $\chi$, we can use the results of \cite{Hamada:2012bp} because the effect of $\chi$ first appears at two-loop level. 
On the other hand, for $X$, $m_{B}^{2}|_{\text{1-loop}}$ becomes
\be \frac{m_{B}^{2}|_{\text{1-loop}}}{\Lambda^{2}/16\pi^{2}}
=-\left(
6\lambda+{3\over4}g_Y^2+{9\over4}g_2^2-6y_t^2+\frac{3}{2}\kappa
\right).
\e
The left (right) panel of Fig.\ref{fig:bare} shows $m_{B}$ as a function of $\Lambda$ when a new particle is fermion (scalar). Here, the green contour is the  
SM prediction when $M_{t}=171.2$ GeV, and blue bands correspond to the $2\sigma$ deviation from it \cite{Moch:2014tta}: 
\be M_{t}=171.2\pm4.8\h{2mm}\text{GeV}\h{3mm}(95\%\text{CL}).\e
In the right panel, we change $\kappa$ at $\mu=M_{X}$ from 0 to 0.4, and they are represented by a red band.
Depending on the values of $M_t$ and $\kappa$, one can see that the scale at which $m_{B}$ becomes zero quite changes. In both of cases, $m_{B}$ can take zero around the string scale \footnote{In order to obtain the correct electroweak symmetry breaking, we need to add small negative mass term to the Higgs potential, which is much small than $\Lambda^{2}$. However, in the case of the $SU(2)_{L}$ triplet scalar, it may be possible to realize  the electroweak symmetry breaking by the Coleman-Weinberg mechanism. See \ref{app:cw}. We thank the referee for pointing this out.}. In addition to the vanishing $\lambda$ at around the string scale,
this fact may suggest the MPP is realized at this scale.\\

%%%%%%%%%%%%%%%%%%%% CONCLUSION %%%%%%%%%%%%%%%%%%%%

In conclusion, we have studied the MPP of the SM with a weakly interacting new particle with its hypercharge being zero. When a new particle is a fermion, we have found that the top mass $M_{t}$ and $ \Lambda_{\text{MPP}}$ can be uniquely predicted. On the other hand, when a new particle is scalar, there exists a new scalar coupling $\kappa$. Due to this coupling, we have found that $ \Lambda_{\text{MPP}}$ and $M_{t}$ drastically change. In both of cases, only the triplets survive from the point of view that the other vacuum should exist around the string/Planck scale and that the theory is valid up to this scale. The analysis of this paper suggests that the SM and  its triplet extensions are special in that the MPP can be realized around the string/Planck scale.

%%%%%%%%%%%%%%%%%%%% ACKNOWLEDGMENTS %%%%%%%%%%%%%%%%%%%%
\section*{Acknowledgement} 
We thank Hikaru Kawai and Koji Tsumura for valuable discussions and useful comments. This work is supported by the Grant-in-Aid for Japan Society for the Promotion of Science (JSPS) Fellows 
No.25$\cdot$1107 (YH) and No.27$\cdot$1771 (KK). 
%%%%%%%%%%%%%%%%%%%% APPENDIX %%%%%%%%%%%%%%%%%%%%

\appendix 
\def\thesection{Appendix \Alph{section}}
\section{Two-loop renormalization group equations and one-loop effective Higgs potential }\label{app:rge}
The two-loop RGEs of the SM with a new fermion which is a $n_{\chi}$ representation of $SU(2)_{L}$ with the hypercharge $Y_{\chi}$ are as follows\footnote{Our calculations are based on \cite{Machacek:1983tz,Machacek:1983fi,Machacek:1984zw,Luo:2002ti}.}:
\begin{align}
\frac{d\Gamma}{dt}&=\frac{1}{(4\pi)^{2}}\left(\frac{9}{4}g_{2}^{2}+\frac{3}{4}g_{Y}^{2}-3y_{t}^{2}\right),
\\
\nonumber\\
\frac{dg_{Y}}{dt}&=\frac{g_{Y}^{3}}{(4\pi)^{2}}\left(\frac{41}{6}+\eta \h{1mm}n_{\chi}\frac{4}{3}Y_{\chi}^{2}\right)+\frac{g_{Y}^{3}}{(4\pi)^{4}}\left\{\left(\frac{199}{18}
   +4\eta\h{1mm}n_{\chi} Y_{\chi}^{4}\right)g_{Y}^2+\left(\frac{9}{2}+4\eta Y_{\chi}^{2}C_{n}\right)g_2^2+\frac{44}{3}
   g_3^2-\frac{17}{6}  y_t^2\right\},\end{align}

\begin{align}
\frac{dg_{2}}{dt}&=\frac{g_{2}^{3}}{(4\pi)^{2}}\left(-\frac{19}{6}+\eta\frac{4}{3}S_{n}\right)+\frac{g_{2}^{3}}{(4\pi)^{4}}\Biggl\{\left(\frac{3}{2}+\eta 4Y_{\chi}^{2}S_{n}\right) g_{Y}^2+\left(\frac{35}{6}+\eta\frac{40}{3}S_{n}+\eta 4C_{n}S_{n}
   \right)g_2^2+12 g_3^2 -\frac{3}{2}
   y_t^2\Biggl\},\nonumber\\
\frac{dg_{3}}{dt}&=-\frac{7}{(4\pi)^{2}}g_{3}^{3}+\frac{g_{3}^{3}}{(4\pi)^{4}}\left(\frac{11}{6}
    g_{Y}^2+\frac{9}{2} g_2^2-26 g_3^2-2  y_t^2
   \right),
\end{align}
\begin{align}
\frac{dy_{t}}{dt}&=\frac{y_{t}}{(4\pi)^{2}}\left(\frac{9}{2}y_{t}^{2}+3y_{\nu}^{2}-8g_{3}^{2}-\frac{9}{4}g_{2}^{2}-\frac{17}{12}g_{Y}^{2}\right)\nonumber
\\
&+\frac{y_{t}}{(4\pi)^{4}}\Biggl\{\left(\frac{1187
   }{216}+\frac{29}{27} \eta \h{1mm}n_{\chi}Y_{\chi}^2\right)g_Y^4+\left(-\frac{23}{4}+\eta S_n\right)g_2^4-\frac{3}{4} g_2^2 g_Y^2+\frac{19}{9} g_3^2 g_Y^2+9 g_3^2 g_2^2-108
   g_3^4\nonumber\\
   &\h{4cm}+\left(\frac{131}{16} g_Y^2 +\frac{225}{16} g_2^2+36 g_3^2
   \right)y_t^2+6 \lambda ^2-12 \lambda  y_t^2-12 y_t^4\Biggl\},
\end{align}

\begin{align}
\frac{d\lambda}{dt}&=\frac{1}{16\pi^{2}}\Biggl(\lambda\left(24
   \lambda-9
   g_{2}^{2}-3  g_{Y}^{2}+12
   y_{t}^{2}\right)+\frac{3}{4}g_{Y}^{2}g_{2}^{2}+\frac{9}{8} g_{2}^{4}+\frac{3}{8}
   g_{Y}^{4}-6 y_{t}^{4}\Biggl)\nonumber
\\  
  &+\frac{1}{(4\pi)^{4}}\Biggl\{-312 \lambda ^3+36\lambda ^2\left(
   g_Y^2+3 g_2^2 \right)+\lambda \left(\frac{629
   }{24}g_Y^4+\frac{10}{3} Y_{\chi}^2 g_Y^4-\frac{73}{8}g_2^4+10 \eta S_ng_2^4 +\frac{39}{4} g_2^2  g_Y^2\right)\nonumber\\
   &+\left(\frac{305}{16}-4
   \eta S_n\right)g_2^6-\left(\frac{289}{48}+\frac{4}{3} \eta S_n\right)g_2^4 g_Y^2-\left(\frac{559}{48}+\frac{4}{3} \eta\h{1mm}n_{\chi} Y_{\chi}^2
   \right)g_2^2 g_Y^4-\left(\frac{379}{48}+\frac{4}{3}n_{\chi}Y_{\chi}^2\right)g_Y^6\nonumber\\
   &+\left(\frac{85}{6} g_Y^2+\frac{45}{2} g_2^2+80 g_3^2\right)\lambda 
   y_t^2+ g_Y^2 y_t^2\left(\frac{21}{2}
   g_2^2-\frac{19}{4} g_Y^2
   \right)-\frac{9}{4} g_2^4 y_t^2-\frac{8}{3} g_Y^2 y_t^4-32 g_3^2 y_t^4\nonumber\\
   &-144 \lambda ^2 y_t^2-3 \lambda
    y_t^4+30 y_t^6
   \Biggl\}.
\end{align}
Here, $t=\ln\mu$ with $\mu$ being the renormalization scale, $\Gamma$ is the wave function renormalization of the Higgs, $C_{n}$ and $S_{n}$ are the Casimir and Dynkin index, and $\eta=1,\frac{1}{2}$ for Dirac and Weyl fermion. 
The two-loop RGEs of $g_Y$ and $g_2$ are agreement with \cite{DiLuzio:2015oha} by putting $y_t=0$.

The one-loop effective Higgs potential is 
\be V_{\text{eff}}(\mu,\phi)=-\frac{M_{h}^{2}}{4}\phi^{2}+\frac{\lambda(\mu)}{4}\phi^{4}+V_{\text{1loop}}(\mu,\phi),\label{eq:Higgs}\e
where
\begin{align} V_{\text{1-loop}}(\mu,\phi):=e^{4\Gamma(\mu)}\Biggl\{-12&\cdot\frac{M_{t}(\phi,\mu)^{4}}{64\pi^{2}}\left[\log\left(\frac{M_{t}(\phi,\mu)^{2}}{\mu^{2}}\right)-\frac{3}{2}+2\Gamma(\mu)\right] \nonumber\\
&+6\cdot\frac{M_{W}(\phi,\mu)^{4}}{64\pi^{2}}\left[\log\left(\frac{M_{W}(\phi,\mu)^{2}}{\mu^{2}}\right)-\frac{5}{6}+2\Gamma(\mu)\right]\nonumber\\
&+3\cdot\frac{M_{Z}(\phi,\mu)^{4}}{64\pi^{2}}\left[\log\left(\frac{M_{Z}(\phi,\mu)^{2}}{\mu^{2}}\right)-\frac{5}{6}+2\Gamma(\mu)\right]\Biggl\},\label{eq:effpot}\end{align}
and
\begin{align} &M_{t}(\phi,\mu)=\frac{y_{t}(\mu)}{\sqrt{2}}\phi\h{2mm},\h{2mm}M_{W}(\phi,\mu)=\frac{g_{2}(\mu)}{2}\phi\h{2mm},\h{2mm}M_{Z}(\phi,\mu)=\frac{\sqrt{g_{2}(\mu)^{2}+g_{Y}(\mu)^{2}}}{2}\phi.\end{align}
In Eq.(\ref{eq:effpot}), we have neglected the contribution from the Higgs quartic term because it is small when we consider the MPP. In principle, $\mu$ should be determined as a function of $\phi$ so that $V_{\text{1-loop}}$ is minimized. However, in this paper, $\mu$ is taken to be $\phi$ for simplicity. It is known that this is a good approximation \cite{Hamada:2014wna}. From $V_{\text{eff}}$, we define $\lambda_{\text{eff}}$ and $\beta_{\lambda_{\text{eff}}}$ as follows:
\be \lambda_{\text{eff}}(\phi):=\frac{4V_{\text{eff}}(\phi)}{\phi^{4}}\h{2mm},\h{2mm}\beta_{\lambda_{\text{eff}}}(\phi):=\frac{d\lambda_{\text{eff}}(\phi)}{d\ln\phi}.\e

%%%%%%%%%%%%%%%%%%%LP%%%%%%%%%%%%%%%%%%%
\section{Landau pole in septet and nonet fermion}\label{app:LP}
\begin{figure}
\begin{center}
\includegraphics[width=12cm]{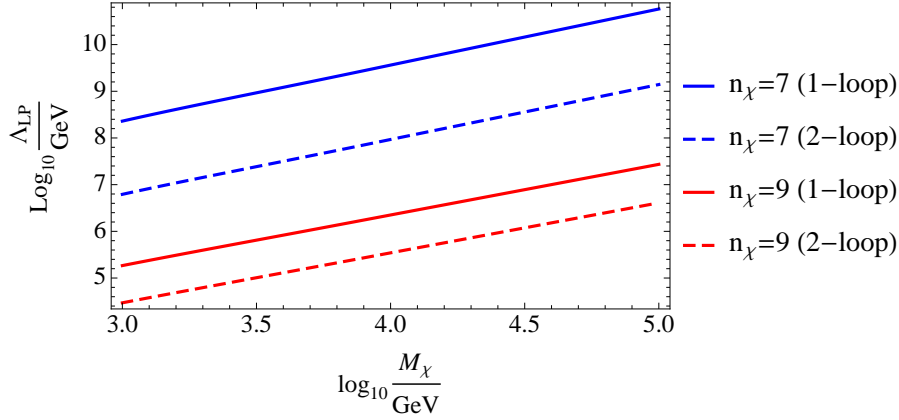}
\end{center}
\caption{The scale $\Lambda_{LP}$ of the Landau pole as a function of $M_{\chi}$ when $n_{\chi}=7$ (blue) and $9$ (red). Here, the two-loop results are represented by dashed lines. }
\label{fig:LP}
\end{figure}

As mentioned in the introduction, in cases of $n_{\chi}=7$ and $9$, there exists a scale $\Lambda_{LP}$ at which $g_{2}$ becomes infinity below $M_{pl}$, which is well known as the Landau Pole. 
Therefore, these theories are not favored from the point of view of perturbativity (triviality) up to the string/Planck scale. 
For completeness, we give numerical results of the Landau pole in Fig.\ref{fig:LP}. Here, the two-loop results are shown by dashed lines. As is known, the one-loop Landau pole can be analytically solved: 
\be \Lambda_{LP}|_{\text{1-loop}}=M_{\chi}\exp\left(\frac{8\pi^{2}}{\left(-\frac{19}{6}+\frac{1}{2}\frac{4}{3}S_{n}\right)g_{2}(M_{\chi})^{2}}\right),\label{eq:lp}\e
where $S_{n}$ is the Dynkin index. From Fig.\ref{fig:LP}, one can see that the two-loop effect is relatively important. 

\section{New Fermion with $Y_{\chi}\neq0$}\label{app:rem}
\begin{figure}[!h]
\begin{center}
\begin{tabular}{c}
\begin{minipage}{0.5\hsize}
\begin{center}
\includegraphics[width=9cm]{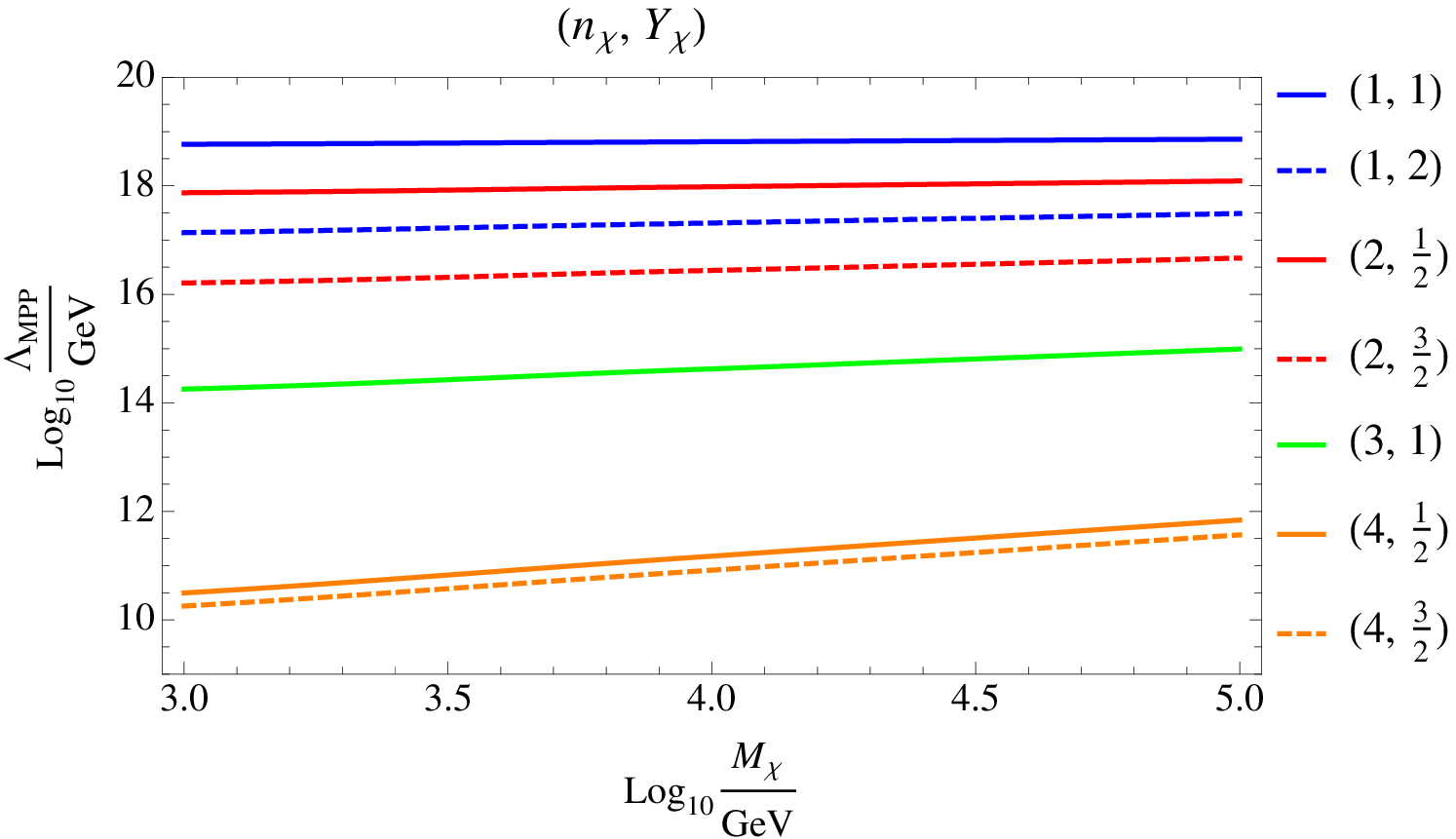}
\end{center}
\end{minipage}
\begin{minipage}{0.5\hsize}
\begin{center}
\includegraphics[width=9cm]{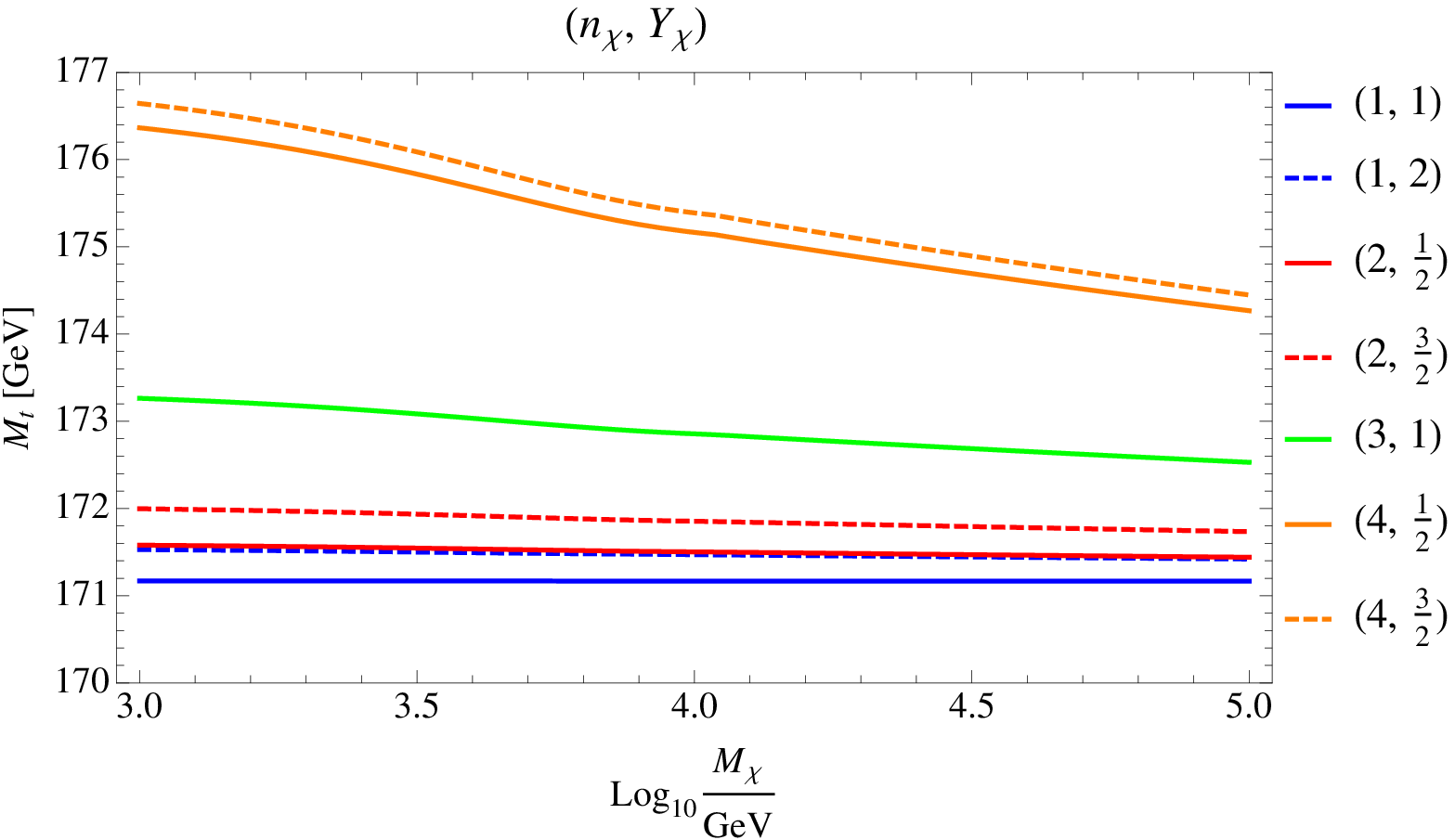}
\end{center}
\end{minipage}
\end{tabular}
\end{center}
\caption{Left (Right): $\Lambda_{\text{MPP}}$ $(M_{t})$ as a function of $M_{\chi}$.}
\label{fig:rem}
\end{figure}

Here, we consider a new fermion with $Y_{\chi}\neq0$. As well as the real $n_{\chi}=7$ and 9 cases, the Landau pole of $g_{2}$ exists below $M_{pl}$ when $n_{\chi}\geq5$ \cite{DiLuzio:2015oha}. So, let us here focus on $n_{\chi}\leq4$ \footnote{For $n_{\chi}=3$ and $4$, the LP of the $U(1)_{Y}$ gauge coupling $g_{Y}$ also appears below $M_{pl}$ respectively when $Y_{\chi}=2$ and $=3/2$. %In the latter case, however, such pole does not cause any problem to the MPP because $\Lambda_{\text{MPP}}\ll\Lambda_{LP}$. 
This is why we only show $Y_{\chi}=1$ when $n_{\chi}=3$ in Fig.\ref{fig:rem}.}. %Because each of them no longer becomes a DM candidate \cite{Cirelli:2009uv}, 
Here, we leave $M_{\chi}$ as a free parameter \footnote{Furthermore, when $n_{\chi}$=1, 2 and 3, there are additional Yukawa couplings among the SM leptons $(L_{i},E_{Ri})$, the Higgs $H$ and $\chi$.
%\be Y^{(1)}_{i}\bar{L^{c}}_{i}H^{\dagger} \chi+\text{h.c}\h{5mm}\text{for $(n_{\chi},Y_{\chi})=(1,1)$},\e
%\be Y^{(2)}_{i}\bar{\chi^{c}} HE_{Ri}+\text{h.c}\h{5mm}\text{for $(n_{\chi},Y_{\chi})=(2,1/2)$},\e
%\be Y^{(3)}_{i}\bar{L^{c}}_{i}\chi H^{\dagger}+\text{h.c}\h{5mm}\text{for $(n_{\chi},Y_{\chi})=(3,1)$},\e
However, we can neglect these effects because the lepton masses are small.}. The left (right) panel of Fig.\ref{fig:rem} shows $\Lambda_{\text{MPP}}$ $(M_{t})$ as a function of $M_{\chi}$ for each $(n_{\chi},Y_{\chi})$. 

\section{Electroweak symmetry breaking by \\Coleman-Weinberg mechanism}\label{app:cw}
Here, we discuss a possibility to realize the electroweak symmetry breaking by the Coleman-Weinberg mechanism in the case of the SU(2) triplet scalar. The one-loop effective Higgs potential is
\aln{ V(\mu,\phi)=\frac{\lambda(\mu)}{4}e^{4\Gamma(\mu)}\phi^{4}&+\frac{3m_{DM}(\phi)^{4}}{64\pi^{2}}\left(\ln\left(\frac{m_{DM}(\phi)^{2}}{\mu^{2}}\right)-\frac{3}{2}\right)\nonumber\\
&+e^{4\Gamma(\mu)}\frac{(3\lambda(\mu)\phi^{2})^{2}}{64\pi^{2}}\left[\log\left(\frac{3\lambda(\mu)e^{2\Gamma(\mu)}\phi^{2}}{\mu^{2}}\right)-\frac{3}{2}\right]+\Delta V_{\text{1-loop}}(\mu,\phi),\label{eq:higgspotentilal}}
where $m_{DM}(\phi)$ and $\Delta V_{\text{1-loop}}(\mu,\phi)$ are given by Eq.(\ref{eq:DMmasspot}) and Eq.(\ref{eq:effpot}) respectively, and we have assumed that the quadratic term vanishes at the tree-level. In the following, we choose $\mu=m_{DM}(\phi)$. Then, $\phi$ develops the vacuum expectation value $v$ because the negative quadratic term appears from the second term in Eq.(\ref{eq:higgspotentilal}). The resultant vacuum expectation value is
\be v=\frac{3M_{X}}{4\pi}\sqrt{\frac{\kappa}{2\lambda}}\simeq 240\text{GeV}\left(\frac{M_{X}}{2.6\text{TeV}}\right)\times \sqrt{\frac{\kappa}{0.04}},\e
where we have neglected the 1-loop correction to the quartic term. It is interesting that the successful electroweak symmetry breaking is realized for $\kappa\simeq0.04$ which is also favored by the MPP around the Planck scale.


\begin{thebibliography}{unsrt}

%\cite{Aad:2012tfa}
\bibitem{Aad:2012tfa} 
  G.~Aad {\it et al.}  [ATLAS Collaboration],
  ``Observation of a new particle in the search for the Standard Model Higgs boson with the ATLAS detector at the LHC,''
  Phys.\ Lett.\ B {\bf 716}, 1 (2012)
  [arXiv:1207.7214 [hep-ex]].
  %%CITATION = ARXIV:1207.7214;%%
  %3465 citations counted in INSPIRE as of 02 Nov 2014

%\cite{Chatrchyan:2012ufa}
\bibitem{Chatrchyan:2012ufa} 
  S.~Chatrchyan {\it et al.}  [CMS Collaboration],
  ``Observation of a new boson at a mass of 125 GeV with the CMS experiment at the LHC,''
  Phys.\ Lett.\ B {\bf 716}, 30 (2012)
  [arXiv:1207.7235 [hep-ex]].
  %%CITATION = ARXIV:1207.7235;%%
  %3404 citations counted in INSPIRE as of 02 Nov 2014

%\cite{Froggatt:1995rt}
\bibitem{Froggatt:1995rt} 
  C.~D.~Froggatt and H.~B.~Nielsen,
  ``Standard model criticality prediction: Top mass 173 +- 5-GeV and Higgs mass 135 +- 9-GeV,''
  Phys.\ Lett.\ B {\bf 368}, 96 (1996)
  [hep-ph/9511371].
  %%CITATION = HEP-PH/9511371;%%
  %126 citations counted in INSPIRE as of 04 Nov 2014

%\cite{Froggatt:2001pa}
\bibitem{Froggatt:2001pa} 
  C.~D.~Froggatt, H.~B.~Nielsen and Y.~Takanishi,
  ``Standard model Higgs boson mass from borderline metastability of the vacuum,''
  Phys.\ Rev.\ D {\bf 64}, 113014 (2001)
  [hep-ph/0104161].
  %%CITATION = HEP-PH/0104161;%%
  %46 citations counted in INSPIRE as of 04 Nov 2014

%\cite{Meissner:2006zh}
\bibitem{Meissner:2006zh} 
  K.~A.~Meissner and H.~Nicolai,
  ``Conformal Symmetry and the Standard Model,''
  Phys.\ Lett.\ B {\bf 648}, 312 (2007)
  [hep-th/0612165].
  %%CITATION = HEP-TH/0612165;%%
  %139 citations counted in INSPIRE as of 04 Nov 2014

%\cite{Foot:2007iy}
\bibitem{Foot:2007iy}
  R.~Foot, A.~Kobakhidze, K.~L.~McDonald and R.~R.~Volkas,
  ``A Solution to the hierarchy problem from an almost decoupled hidden sector within a classically scale invariant theory,''
  Phys.\ Rev.\ D {\bf 77} (2008) 035006
  [arXiv:0709.2750 [hep-ph]].
  %%CITATION = ARXIV:0709.2750;%%
  %110 citations counted in INSPIRE as of 21 juin 2015


%\cite{Meissner:2007xv}
\bibitem{Meissner:2007xv} 
  K.~A.~Meissner and H.~Nicolai,
  ``Effective action, conformal anomaly and the issue of quadratic divergences,''
  Phys.\ Lett.\ B {\bf 660}, 260 (2008)
  [arXiv:0710.2840 [hep-th]].
  %%CITATION = ARXIV:0710.2840;%%
  %59 citations counted in INSPIRE as of 04 Nov 2014

%\cite{Iso:2009ss}
\bibitem{Iso:2009ss}
  S.~Iso, N.~Okada and Y.~Orikasa,
  ``Classically conformal $B^-$ L extended Standard Model,''
  Phys.\ Lett.\ B {\bf 676} (2009) 81
  [arXiv:0902.4050 [hep-ph]].
  %%CITATION = ARXIV:0902.4050;%%
  %112 citations counted in INSPIRE as of 21 juin 2015

%\cite{Iso:2009nw}
\bibitem{Iso:2009nw}
  S.~Iso, N.~Okada and Y.~Orikasa,
  ``The minimal B-L model naturally realized at TeV scale,''
  Phys.\ Rev.\ D {\bf 80} (2009) 115007
  [arXiv:0909.0128 [hep-ph]].
  %%CITATION = ARXIV:0909.0128;%%
  %80 citations counted in INSPIRE as of 21 juin 2015




%\cite{Shaposhnikov:2009pv}
\bibitem{Shaposhnikov:2009pv} 
  M.~Shaposhnikov and C.~Wetterich,
  ``Asymptotic safety of gravity and the Higgs boson mass,''
  Phys.\ Lett.\ B {\bf 683}, 196 (2010)
  [arXiv:0912.0208 [hep-th]].
  %%CITATION = ARXIV:0912.0208;%%
  %115 citations counted in INSPIRE as of 04 Nov 2014

%\cite{Holthausen:2011aa}
\bibitem{Holthausen:2011aa} 
  M.~Holthausen, K.~S.~Lim and M.~Lindner,
  ``Planck scale Boundary Conditions and the Higgs Mass,''
  JHEP {\bf 1202}, 037 (2012)
  [arXiv:1112.2415 [hep-ph]].
  %%CITATION = ARXIV:1112.2415;%%
  %128 citations counted in INSPIRE as of 12 juin 2015

%\cite{Bezrukov:2012sa}
\bibitem{Bezrukov:2012sa} 
  F.~Bezrukov, M.~Y.~Kalmykov, B.~A.~Kniehl and M.~Shaposhnikov,
  ``Higgs Boson Mass and New Physics,''
  JHEP {\bf 1210}, 140 (2012)
  [arXiv:1205.2893 [hep-ph]].
  %%CITATION = ARXIV:1205.2893;%%
  %234 citations counted in INSPIRE as of 12 juin 2015

  %\cite{Hamada:2012bp}
\bibitem{Hamada:2012bp} 
  Y.~Hamada, H.~Kawai and K.~y.~Oda,
  ``Bare Higgs mass at Planck scale,''
  Phys.\ Rev.\ D {\bf 87}, no. 5, 053009 (2013)
  [Phys.\ Rev.\ D {\bf 89}, no. 5, 059901 (2014)]
  [arXiv:1210.2538 [hep-ph]].
  %%CITATION = ARXIV:1210.2538;%%
  %46 citations counted in INSPIRE as of 24 Apr 2015

%\cite{Iso:2012jn}
\bibitem{Iso:2012jn}
  S.~Iso and Y.~Orikasa,
  ``TeV Scale B-L model with a flat Higgs potential at the Planck scale - in view of the hierarchy problem -,''
  PTEP {\bf 2013} (2013) 023B08
  [arXiv:1210.2848 [hep-ph]].
  %%CITATION = ARXIV:1210.2848;%%
  %71 citations counted in INSPIRE as of 21 juin 2015


%\cite{Nielsen:2012pu}
\bibitem{Nielsen:2012pu} 
  H.~B.~Nielsen,
  ``PREdicted the Higgs Mass,''
  arXiv:1212.5716 [hep-ph].
  %%CITATION = ARXIV:1212.5716;%%
  %12 citations counted in INSPIRE as of 04 Nov 2014

%\cite{Jegerlehner:2013cta}
\bibitem{Jegerlehner:2013cta}
  F.~Jegerlehner,
  ``The Standard model as a low-energy effective theory: what is triggering the Higgs mechanism?,''
  Acta Phys.\ Polon.\ B {\bf 45} (2014) 6,  1167
  [arXiv:1304.7813 [hep-ph]].
  %%CITATION = ARXIV:1304.7813;%%
  %33 citations counted in INSPIRE as of 21 juin 2015

%\cite{Jegerlehner:2013nna}
\bibitem{Jegerlehner:2013nna}
  F.~Jegerlehner,
  ``The hierarchy problem of the electroweak Standard Model revisited,''
  arXiv:1305.6652 [hep-ph].
  %%CITATION = ARXIV:1305.6652;%%
  %22 citations counted in INSPIRE as of 21 juin 2015


%\cite{Hamada:2013cta}
\bibitem{Hamada:2013cta}
  Y.~Hamada, H.~Kawai and K.~y.~Oda,
  ``Bare Higgs mass and potential at ultraviolet cutoff,''
  arXiv:1305.7055 [hep-ph].
  %%CITATION = ARXIV:1305.7055;%%
  %12 citations counted in INSPIRE as of 21 Jun 2015

%\cite{Buttazzo:2013uya}
\bibitem{Buttazzo:2013uya} 
  D.~Buttazzo, G.~Degrassi, P.~P.~Giardino, G.~F.~Giudice, F.~Sala, A.~Salvio and A.~Strumia,
  ``Investigating the near-criticality of the Higgs boson,''
  JHEP {\bf 1312}, 089 (2013)
  [arXiv:1307.3536 [hep-ph]].
  %%CITATION = ARXIV:1307.3536;%%
  %172 citations counted in INSPIRE as of 29 Oct 2014

%\cite{Branchina:2013jra}
\bibitem{Branchina:2013jra}
  V.~Branchina and E.~Messina,
  %``Stability, Higgs Boson Mass and New Physics,''
  Phys.\ Rev.\ Lett.\  {\bf 111} (2013) 241801
  [arXiv:1307.5193 [hep-ph]].
  %%CITATION = ARXIV:1307.5193;%%
  %63 citations counted in INSPIRE as of 21 Jun 2015

%\cite{Kawamura:2013kua}
\bibitem{Kawamura:2013kua} 
  Y.~Kawamura,
  ``Naturalness, Conformal Symmetry and Duality,''
  PTEP {\bf 2013}, no. 11, 113B04 (2013)
  [arXiv:1308.5069 [hep-ph]].
  %%CITATION = ARXIV:1308.5069;%%
  %15 citations counted in INSPIRE as of 04 Nov 2014

%\cite{Chao:2012mx}
\bibitem{Chao:2012mx} 
  W.~Chao, M.~Gonderinger and M.~J.~Ramsey-Musolf,
  ``Higgs Vacuum Stability, Neutrino Mass, and Dark Matter,''
  Phys.\ Rev.\ D {\bf 86}, 113017 (2012)
  [arXiv:1210.0491 [hep-ph]].
  %%CITATION = ARXIV:1210.0491;%%
  %29 citations counted in INSPIRE as of 28 Jul 2015

%\cite{Kobakhidze:2014xda}
\bibitem{Kobakhidze:2014xda}
  A.~Kobakhidze and A.~Spencer-Smith,
  ``The Higgs vacuum is unstable,''
  arXiv:1404.4709 [hep-ph].
  %%CITATION = ARXIV:1404.4709;%%
  %33 citations counted in INSPIRE as of 21 juin 2015

%\cite{Khan:2014kba}
\bibitem{Khan:2014kba} 
  N.~Khan and S.~Rakshit,
  ``Study of electroweak vacuum metastability with a singlet scalar dark matter,''
  Phys.\ Rev.\ D {\bf 90}, no. 11, 113008 (2014)
  [arXiv:1407.6015 [hep-ph]].
  %%CITATION = ARXIV:1407.6015;%%
  %6 citations counted in INSPIRE as of 29 juil. 2015

%\cite{Khan:2015ipa}
\bibitem{Khan:2015ipa} 
  N.~Khan and S.~Rakshit,
  ``Constraints on inert dark matter from metastability of electroweak vacuum,''
  arXiv:1503.03085 [hep-ph].
  %%CITATION = ARXIV:1503.03085;%%
  %6 citations counted in INSPIRE as of 29 juil. 2015

%\cite{Spencer-Smith:2014woa}
\bibitem{Spencer-Smith:2014woa}
  A.~Spencer-Smith,
  ``Higgs Vacuum Stability in a Mass-Dependent Renormalisation Scheme,''
  arXiv:1405.1975 [hep-ph].
  %%CITATION = ARXIV:1405.1975;%%
  %25 citations counted in INSPIRE as of 21 juin 2015

%\cite{Haba:2014sia}
\bibitem{Haba:2014sia} 
  N.~Haba, H.~Ishida, K.~Kaneta and R.~Takahashi,
  ``Vanishing Higgs potential at the Planck scale in a singlet extension of the standard model,''
  Phys.\ Rev.\ D {\bf 90}, 036006 (2014)
  [arXiv:1406.0158 [hep-ph]].
  %%CITATION = ARXIV:1406.0158;%%
  %3 citations counted in INSPIRE as of 07 Nov 2014


%\cite{Foot:2014ifa}
\bibitem{Foot:2014ifa}
  R.~Foot, A.~Kobakhidze and A.~Spencer-Smith,
  ``Criticality in the scale invariant standard model (squared),''
  Phys.\ Lett.\ B {\bf 747} (2015) 169
  [arXiv:1409.4915 [hep-ph]].
  %%CITATION = ARXIV:1409.4915;%%
  %2 citations counted in INSPIRE as of 21 juin 2015

%\cite{Oda:2015kma}
\bibitem{Oda:2015kma}
  I.~Oda,
  ``Conformal Higgs Gravity,''
  arXiv:1505.06760 [gr-qc].
  %%CITATION = ARXIV:1505.06760;%%


%\cite{Bezrukov:2007ep}
\bibitem{Bezrukov:2007ep} 
  F.~L.~Bezrukov and M.~Shaposhnikov,
  ``The Standard Model Higgs boson as the inflaton,''
  Phys.\ Lett.\ B {\bf 659}, 703 (2008)
  [arXiv:0710.3755 [hep-th]].
  %%CITATION = ARXIV:0710.3755;%%
  %522 citations counted in INSPIRE as of 26 Dec 2014


%\cite{Hamada:2013mya}
\bibitem{Hamada:2013mya} 
  Y.~Hamada, H.~Kawai and K.~y.~Oda,
  ``Minimal Higgs inflation,''
  PTEP {\bf 2014}, 023B02 (2014)
  [arXiv:1308.6651 [hep-ph]].
  %%CITATION = ARXIV:1308.6651;%%
  %18 citations counted in INSPIRE as of 06 Nov 2014

%\cite{Jegerlehner:2014mua}
\bibitem{Jegerlehner:2014mua}
  F.~Jegerlehner,
  ``Higgs inflation and the cosmological constant,''
  Acta Phys.\ Polon.\ B {\bf 45} (2014) 6,  1215
  [arXiv:1402.3738 [hep-ph]].
  %%CITATION = ARXIV:1402.3738;%%
  %8 citations counted in INSPIRE as of 21 juin 2015

%\cite{Hamada:2014iga}
\bibitem{Hamada:2014iga} 
  Y.~Hamada, H.~Kawai, K.~y.~Oda and S.~C.~Park,
  ``Higgs inflation still alive,''
  Phys.\ Rev.\ Lett.\  {\bf 112}, 241301 (2014)
  [arXiv:1403.5043 [hep-ph]].
  %%CITATION = ARXIV:1403.5043;%%
  %46 citations counted in INSPIRE as of 06 Nov 2014

%\cite{Fairbairn:2014zia}
\bibitem{Fairbairn:2014zia} 
  M.~Fairbairn and R.~Hogan,
  ``Electroweak Vacuum Stability in light of BICEP2,''
  Phys.\ Rev.\ Lett.\  {\bf 112}, 201801 (2014)
  [arXiv:1403.6786 [hep-ph]].
  %%CITATION = ARXIV:1403.6786;%%
  %35 citations counted in INSPIRE as of 29 juil. 2015

%\cite{Bezrukov:2014bra}
\bibitem{Bezrukov:2014bra}
  F.~Bezrukov and M.~Shaposhnikov,
  ``Higgs inflation at the critical point,''
  Phys.\ Lett.\ B {\bf 734} (2014) 249
  [arXiv:1403.6078 [hep-ph]].
  %%CITATION = ARXIV:1403.6078;%%
  %54 citations counted in INSPIRE as of 21 Jun 2015

%\cite{Enqvist:2014bua}
\bibitem{Enqvist:2014bua}
  K.~Enqvist, T.~Meriniemi and S.~Nurmi,
  ``Higgs Dynamics during Inflation,''
  JCAP {\bf 1407} (2014) 025
  [arXiv:1404.3699 [hep-ph]].
  %%CITATION = ARXIV:1404.3699;%%
  %25 citations counted in INSPIRE as of 21 juin 2015

%\cite{Hook:2014uia}
\bibitem{Hook:2014uia} 
  A.~Hook, J.~Kearney, B.~Shakya and K.~M.~Zurek,
  ``Probable or Improbable Universe? Correlating Electroweak Vacuum Instability with the Scale of Inflation,''
  JHEP {\bf 1501}, 061 (2015)
  [arXiv:1404.5953 [hep-ph]].
  %%CITATION = ARXIV:1404.5953;%%
  %25 citations counted in INSPIRE as of 29 juil. 2015

%\cite{Haba:2014zda}
\bibitem{Haba:2014zda}
  N.~Haba and R.~Takahashi,
  ``Higgs inflation with singlet scalar dark matter and right-handed neutrino in light of BICEP2,''
  Phys.\ Rev.\ D {\bf 89} (2014) 11,  115009
   [Phys.\ Rev.\ D {\bf 90} (2014) 3,  039905]
  [arXiv:1404.4737 [hep-ph]].
  %%CITATION = ARXIV:1404.4737;%%
  %24 citations counted in INSPIRE as of 21 juin 2015


%\cite{Hamada:2014xka}
\bibitem{Hamada:2014xka} 
  Y.~Hamada, H.~Kawai and K.~y.~Oda,
  ``Predictions on mass of Higgs portal scalar dark matter from Higgs inflation and flat potential,''
  JHEP {\bf 1407}, 026 (2014)
  [arXiv:1404.6141 [hep-ph]].
  %%CITATION = ARXIV:1404.6141;%%
  %14 citations counted in INSPIRE as of 27 Oct 2014

%\cite{Ko:2014eia}
\bibitem{Ko:2014eia}
  P.~Ko and W.~I.~Park,
  ``Higgs-portal assisted Higgs inflation with a large tensor-to-scalar ratio,''
  arXiv:1405.1635 [hep-ph].
  %%CITATION = ARXIV:1405.1635;%%
  %16 citations counted in INSPIRE as of 21 juin 2015


%\cite{Haba:2014zja}
\bibitem{Haba:2014zja}
  N.~Haba, H.~Ishida and R.~Takahashi,
  ``Higgs inflation and Higgs portal dark matter with right-handed neutrinos,''
  PTEP {\bf 2015} (2015) 5,  053B01
  [arXiv:1405.5738 [hep-ph]].
  %%CITATION = ARXIV:1405.5738;%%
  %12 citations counted in INSPIRE as of 21 juin 2015

%\cite{He:2014ora}
\bibitem{He:2014ora}
  H.~J.~He and Z.~Z.~Xianyu,
  ``Extending Higgs Inflation with TeV Scale New Physics,''
  JCAP {\bf 1410} (2014) 019
  [arXiv:1405.7331 [hep-ph]].
  %%CITATION = ARXIV:1405.7331;%%
  %12 citations counted in INSPIRE as of 21 juin 2015

%\cite{Herranen:2014cua}
\bibitem{Herranen:2014cua} 
  M.~Herranen, T.~Markkanen, S.~Nurmi and A.~Rajantie,
  ``Spacetime curvature and the Higgs stability during inflation,''
  Phys.\ Rev.\ Lett.\  {\bf 113}, no. 21, 211102 (2014)
  [arXiv:1407.3141 [hep-ph]].
  %%CITATION = ARXIV:1407.3141;%%
  %31 citations counted in INSPIRE as of 29 juil. 2015

%\cite{Hamada:2014wna}
\bibitem{Hamada:2014wna}
  Y.~Hamada, H.~Kawai, K.~y.~Oda and S.~C.~Park,
  ``Higgs inflation from Standard Model criticality,''
  Phys.\ Rev.\ D {\bf 91} (2015) 5,  053008
  [arXiv:1408.4864 [hep-ph]].
  %%CITATION = ARXIV:1408.4864;%%
  %16 citations counted in INSPIRE as of 19 May 2015

%\cite{Hamada:2014raa}
\bibitem{Hamada:2014raa}
  Y.~Hamada, K.~y.~Oda and F.~Takahashi,
  ``Topological Higgs inflation: Origin of Standard Model criticality,''
  Phys.\ Rev.\ D {\bf 90} (2014) 9,  097301
  [arXiv:1408.5556 [hep-ph]].
  %%CITATION = ARXIV:1408.5556;%%
  %12 citations counted in INSPIRE as of 21 Jun 2015

%\cite{Hamada:2015ria}
\bibitem{Hamada:2015ria} 
  Y.~Hamada, H.~Kawai and K.~y.~Oda,
  ``Eternal Higgs inflation and cosmological constant problem,''
  arXiv:1501.04455 [hep-ph].
  %%CITATION = ARXIV:1501.04455;%%
  %1 citations counted in INSPIRE as of 13 Feb 2015

%\cite{Okada:2015zfa}
\bibitem{Okada:2015zfa} 
  N.~Okada and Q.~Shafi,
  ``Higgs Inflation, Seesaw Physics and Fermion Dark Matter,''
  Phys.\ Lett.\ B {\bf 747}, 223 (2015)
  [arXiv:1501.05375 [hep-ph]].
  %%CITATION = ARXIV:1501.05375;%%
  %1 citations counted in INSPIRE as of 22 Jun 2015


%\cite{Inagaki:2015fva}
\bibitem{Inagaki:2015fva}
  T.~Inagaki, R.~Nakanishi and S.~D.~Odintsov,
  ``Non-Minimal Two-Loop Inflation,''
  Phys.\ Lett.\ B {\bf 745} (2015) 105
  [arXiv:1502.06301 [hep-ph]].
  %%CITATION = ARXIV:1502.06301;%%
  %4 citations counted in INSPIRE as of 21 juin 2015



%\cite{Jegerlehner:2015cva}
\bibitem{Jegerlehner:2015cva}
  F.~Jegerlehner,
  ``The hierarchy problem and the cosmological constant problem in the Standard Model,''
  arXiv:1503.00809 [hep-ph].
  %%CITATION = ARXIV:1503.00809;%%

%\cite{Abe:2015bba}
\bibitem{Abe:2015bba}
  Y.~Abe, T.~Inami, Y.~Kawamura and Y.~Koyama,
  ``Inflation from radion gauge-Higgs potential at Planck scale,''
  arXiv:1504.06905 [hep-th].
  %%CITATION = ARXIV:1504.06905;%%
  %1 citations counted in INSPIRE as of 21 juin 2015

\bibitem{DiLuzio:2015oha}
  L.~Di Luzio, R.~Grober, J.~F.~Kamenik and M.~Nardecchia,
  ``Accidental matter at the LHC,''
  arXiv:1504.00359 [hep-ph]. 


%\cite{Bamba:2015uxa}
\bibitem{Bamba:2015uxa}
  K.~Bamba, S.~D.~Odintsov and P.~V.~Tretyakov,
  ``Inflation in a conformally-invariant two-scalar-field theory with an extra $R^2$ term,''
  arXiv:1505.00854 [hep-th].
  %%CITATION = ARXIV:1505.00854;%%

%\cite{Nurmi:2015ema}
\bibitem{Nurmi:2015ema}
  S.~Nurmi, T.~Tenkanen and K.~Tuominen,
  ``Inflationary Imprints on Dark Matter,''
  arXiv:1506.04048 [astro-ph.CO].
  %%CITATION = ARXIV:1506.04048;%%

%\cite{Sebastiani:2015kfa}
\bibitem{Sebastiani:2015kfa}
  L.~Sebastiani and R.~Myrzakulov,
  ``F(R) gravity and inflation,''
  arXiv:1506.05330 [gr-qc].
  %%CITATION = ARXIV:1506.05330;%%

%\cite{Herranen:2015ima}
\bibitem{Herranen:2015ima} 
  M.~Herranen, T.~Markkanen, S.~Nurmi and A.~Rajantie,
  ``Spacetime curvature and Higgs stability after inflation,''
  arXiv:1506.04065 [hep-ph].
  %%CITATION = ARXIV:1506.04065;%%
  %2 citations counted in INSPIRE as of 29 juil. 2015

%\cite{Kawai:2011qb}
\bibitem{Kawai:2011qb} 
  H.~Kawai and T.~Okada,
  ``Solving the Naturalness Problem by Baby Universes in the Lorentzian Multiverse,''
  Prog.\ Theor.\ Phys.\  {\bf 127}, 689 (2012)
  [arXiv:1110.2303 [hep-th]].
  %%CITATION = ARXIV:1110.2303;%%
  %14 citations counted in INSPIRE as of 04 Nov 2014

%\cite{Kawai:2013wwa}
\bibitem{Kawai:2013wwa} 
  H.~Kawai,
  ``Low energy effective action of quantum gravity and the naturalness problem,''
  Int.\ J.\ Mod.\ Phys.\ A {\bf 28}, 1340001 (2013).
  %%CITATION = IMPAE,A28,1340001;%%
  %6 citations counted in INSPIRE as of 04 Nov 2014

%\cite{Hamada:2014ofa}
\bibitem{Hamada:2014ofa} 
  Y.~Hamada, H.~Kawai and K.~Kawana,
  ``Evidence of the Big Fix,''
  Int.\ J.\ Mod.\ Phys.\ A {\bf 29}, no. 17, 1450099 (2014)
  [arXiv:1405.1310 [hep-ph]].
  %%CITATION = ARXIV:1405.1310;%%
  %4 citations counted in INSPIRE as of 04 Nov 2014

%\cite{Hamada:2014xra}
\bibitem{Hamada:2014xra}
  Y.~Hamada, H.~Kawai and K.~Kawana,
  ``Weak Scale From the Maximum Entropy Principle,''
  PTEP {\bf 2015} (2015) 3,  033B06
  [arXiv:1409.6508 [hep-ph]].
  %%CITATION = ARXIV:1409.6508;%%
  %6 citations counted in INSPIRE as of 21 May 2015
  
  %\cite{Hamada:2015wea}
\bibitem{Hamada:2015wea} 
  Y.~Hamada, H.~Kawai and K.~Kawana,
  ``Saddle point inflation in string-inspired theory,''
  PTEP {\bf 2015}, 091B01 (2015)
  [arXiv:1507.03106 [hep-ph]].
  %%CITATION = ARXIV:1507.03106;%%
  %3 citations counted in INSPIRE as of 03 Oct 2015
  
  %\cite{Hamada:2015dja}
\bibitem{Hamada:2015dja} 
  Y.~Hamada, H.~Kawai and K.~Kawana,
  ``Natural solution to the naturalness problem -- Universe does fine-tuning,''
  arXiv:1509.05955 [hep-th].
  %%CITATION = ARXIV:1509.05955;%%

%\cite{Kawana:2014zxa}
\bibitem{Kawana:2014zxa} 
  K.~Kawana,
  ``Multiple Point Principle of the Standard Model with Scalar Singlet Dark Matter and Right Handed Neutrinos,''
  PTEP {\bf 2015}, no. 2, 023B04
  [arXiv:1411.2097 [hep-ph]].
  %%CITATION = ARXIV:1411.2097;%%
  %8 citations counted in INSPIRE as of 24 Apr 2015

%\cite{Okada:2014nea}
\bibitem{Okada:2014nea}
  H.~Okada and Y.~Orikasa,
  ``Classically Conformal Radiative Neutrino Model with Gauged B-L Symmetry,''
  arXiv:1412.3616 [hep-ph].
  %%CITATION = ARXIV:1412.3616;%%
  %11 citations counted in INSPIRE as of 21 juin 2015

%\cite{Kawana:2015tka}
\bibitem{Kawana:2015tka} 
  K.~Kawana,
  ``Criticality and Inflation of the Gauged B-L Model,''
  arXiv:1501.04482 [hep-ph].
  %%CITATION = ARXIV:1501.04482;%%
  %6 citations counted in INSPIRE as of 24 Apr 2015

%\cite{Haba:2015rha}
\bibitem{Haba:2015rha}
  N.~Haba and Y.~Yamaguchi,
  ``Vacuum stability in the $U(1)_\chi$ extended model with vanishing scalar potential at the Planck scale,''
  arXiv:1504.05669 [hep-ph].
  %%CITATION = ARXIV:1504.05669;%%

%\cite{Cirelli:2005uq}
\bibitem{Cirelli:2005uq} 
  M.~Cirelli, N.~Fornengo and A.~Strumia,
  ``Minimal dark matter,''
  Nucl.\ Phys.\ B {\bf 753}, 178 (2006)
  [hep-ph/0512090].
  %%CITATION = HEP-PH/0512090;%%
  %331 citations counted in INSPIRE as of 24 Apr 2015

%\cite{Hisano:2006nn}
\bibitem{Hisano:2006nn} 
  J.~Hisano, S.~Matsumoto, M.~Nagai, O.~Saito and M.~Senami,
  ``Non-perturbative effect on thermal relic abundance of dark matter,''
  Phys.\ Lett.\ B {\bf 646}, 34 (2007)
  [hep-ph/0610249].
  %%CITATION = HEP-PH/0610249;%%
  %184 citations counted in INSPIRE as of 15 Jun 2015

%\cite{Cirelli:2007xd}
\bibitem{Cirelli:2007xd} 
  M.~Cirelli, A.~Strumia and M.~Tamburini,
  ``Cosmology and Astrophysics of Minimal Dark Matter,''
  Nucl.\ Phys.\ B {\bf 787}, 152 (2007)
  [arXiv:0706.4071 [hep-ph]].
  %%CITATION = ARXIV:0706.4071;%%
  %231 citations counted in INSPIRE as of 24 Apr 2015

%\cite{Cirelli:2009uv}
%\bibitem{Cirelli:2009uv} 
  %M.~Cirelli and A.~Strumia,
  %``Minimal Dark Matter: Model and results,''
  %New J.\ Phys.\  {\bf 11}, 105005 (2009)
  %[arXiv:0903.3381 [hep-ph]].
  %%CITATION = ARXIV:0903.3381;%%
  %71 citations counted in INSPIRE as of 24 Apr 2015



%\cite{Hamada:2015bra}
\bibitem{Hamada:2015bra} 
  Y.~Hamada, K.~Kawana and K.~Tsumura,
  ``Landau pole in the Standard Model with weakly interacting scalar fields,''
  Phys.\ Lett.\ B {\bf 747}, 238 (2015)
  [arXiv:1505.01721 [hep-ph]].
  %%CITATION = ARXIV:1505.01721;%%

%\cite{Aad:2015zhl}
\bibitem{Aad:2015zhl} 
  G.~Aad {\it et al.} [ATLAS and CMS Collaborations],
  ``Combined Measurement of the Higgs Boson Mass in $pp$ Collisions at $\sqrt{s}=7$ and 8 TeV with the ATLAS and CMS Experiments,''
  Phys.\ Rev.\ Lett.\  {\bf 114}, 191803 (2015)
  [arXiv:1503.07589 [hep-ex]].
  %%CITATION = ARXIV:1503.07589;%%
  %94 citations counted in INSPIRE as of 25 juil. 2015


  %\cite{ATLAS:2014wva}
\bibitem{ATLAS:2014wva} 
  [ATLAS and CDF and CMS and D0 Collaborations],
  ``First combination of Tevatron and LHC measurements of the top-quark mass,''
  arXiv:1403.4427 [hep-ex].
  %%CITATION = ARXIV:1403.4427;%%
  %235 citations counted in INSPIRE as of 03 Oct 2015

%\cite{CMS:2014hta}
\bibitem{CMS:2014hta} 
  CMS Collaboration [CMS Collaboration],
  ``Combination of the CMS top-quark mass measurements from Run 1 of the LHC,''
  CMS-PAS-TOP-14-015.
  %%CITATION = CMS-PAS-TOP-14-015;%%
  %11 citations counted in INSPIRE as of 03 Oct 2015

  %\cite{Moch:2014tta}
\bibitem{Moch:2014tta} 
  S.~Moch, S.~Weinzierl, S.~Alekhin, J.~Blumlein, L.~de la Cruz, S.~Dittmaier, M.~Dowling and J.~Erler {\it et al.},
  ``High precision fundamental constants at the TeV scale,''
  arXiv:1405.4781 [hep-ph].
  %%CITATION = ARXIV:1405.4781;%%
  %40 citations counted in INSPIRE as of 24 Apr 2015

%\cite{Veltman:1980mj}
\bibitem{Veltman:1980mj}
  M.~J.~G.~Veltman,
  ``The Infrared - Ultraviolet Connection,''
  Acta Phys.\ Polon.\ B {\bf 12} (1981) 437.
  %%CITATION = APPOA,B12,437;%%
  %542 citations counted in INSPIRE as of 19 May 2015

 

%\cite{Machacek:1983tz}
\bibitem{Machacek:1983tz} 
  M.~E.~Machacek and M.~T.~Vaughn,
  ``Two Loop Renormalization Group Equations in a General Quantum Field Theory. 1. Wave Function Renormalization,''
  Nucl.\ Phys.\ B {\bf 222}, 83 (1983).
  %%CITATION = NUPHA,B222,83;%%
  %423 citations counted in INSPIRE as of 01 Nov 2014

%\cite{Machacek:1983fi}
\bibitem{Machacek:1983fi} 
  M.~E.~Machacek and M.~T.~Vaughn,
  ``Two Loop Renormalization Group Equations in a General Quantum Field Theory. 2. Yukawa Couplings,''
  Nucl.\ Phys.\ B {\bf 236}, 221 (1984).
  %%CITATION = NUPHA,B236,221;%%
  %380 citations counted in INSPIRE as of 01 Nov 2014
  
%\cite{Machacek:1984zw}
\bibitem{Machacek:1984zw} 
  M.~E.~Machacek and M.~T.~Vaughn,
  ``Two Loop Renormalization Group Equations in a General Quantum Field Theory. 3. Scalar Quartic Couplings,''
  Nucl.\ Phys.\ B {\bf 249}, 70 (1985).
  %%CITATION = NUPHA,B249,70;%%
  %304 citations counted in INSPIRE as of 01 Nov 2014

%\cite{Luo:2002ti}
\bibitem{Luo:2002ti} 
  M.~x.~Luo, H.~w.~Wang and Y.~Xiao,
  ``Two loop renormalization group equations in general gauge field theories,''
  Phys.\ Rev.\ D {\bf 67}, 065019 (2003)
  [hep-ph/0211440].
  %%CITATION = HEP-PH/0211440;%%
  %59 citations counted in INSPIRE as of 24 Apr 2015
  


\end{thebibliography}
\end{document}